\documentclass{ar-1col-S2O}

\usepackage[comma]{natbib}
\usepackage{url}

\usepackage{amssymb}
\usepackage{hyperref}
\hypersetup{
    colorlinks=true,
    linkcolor=blue,
    filecolor=magenta,      
    urlcolor=cyan,
    citecolor=blue,
    linktoc=page,
    pdftitle={Overleaf Example},
    pdfpagemode=FullScreen,
}

\jname{Xxxx. Xxx. Xxx. Xxx.}
\jvol{AA}
\jyear{YYYY}
\doi{10.1146/((please add article doi))}

\begin{document}

\markboth{Huber}{Space-Based Time-Domain Astronomy}

\title{The Space-Based Time-Domain Revolution in Astrophysics}

\author{Daniel Huber,$^{1,2}$ 
\affil{$^1$Institute for Astronomy, University of Hawai`i, 2680 Woodlawn Drive, Honolulu, HI 96822, USA ; email: huberd@hawaii.edu}
\affil{$^2$Sydney Institute for Astronomy (SIfA), School of Physics, University of Sydney, NSW 2006, Australia}}
\begin{abstract}

Space-based time-domain telescopes such as CoRoT, Kepler/K2 and TESS have profoundly impacted astrophysics over the past two decades. Continuous light curves with high cadence and high photometric precision are now available for millions of sources within our galaxy and beyond. In addition to revolutionizing exoplanet science, the data have enabled breakthroughs ranging from the solar system to stellar interiors, the transient universe, and active galaxies. 

\hangindent=.3cm$\bullet$ Stellar astrophysics has been transformed by the ability to probe the internal structures of stars, test the physics of stellar convection, connect stellar rotation and magnetic activity, and reveal complex variability in young stars.

\hangindent=.3cm$\bullet$ Ages of stellar populations probe the formation history of our Milky Way, and binary star variability enables the detection of ``dark'' galactic populations such as stellar-mass black holes and neutron stars.

\hangindent=.3cm$\bullet$  Early-time observations of explosive transients provide new insights into the progenitors of supernovae, while the quasi-periodic variability of galaxies probes the physics of accretion processes onto supermassive black holes and the tidal disruption of stars.

\hangindent=.3cm$\bullet$  Observations of solar system objects reveal asteroid compositions through their rotation periods and amplitudes, constrain the cloud structure of ice giants, and allow the discovery of new objects in the outer solar system.

\hangindent=.3cm$\bullet$ Open data policies and software have contributed to remarkable scientific productivity and enabled discoveries by citizen scientists, including new exoplanets and exotic variability in mature Sun-like stars.

\end{abstract}

\maketitle

\tableofcontents

\section{Introduction}

\subsection{Scope of this Review} 

\noindent
Time-domain photometry --- measuring the brightness of objects over multiple epochs --- has a long history in astronomy. Examples include the first recorded supernova explosions by Chinese astronomers over 2000 years ago \citep[``guest stars'',][]{clark_historical_1977}, the naked-eye detection of variability in Mira by David Fabricius in the 16th century \citep{olbers_materialien_1850}, the discovery of eclipsing binary stars \citep{goodricke_series_1783} and the pioneering observations of Cepheid pulsations that led to the discovery of the expansion of the Universe \citep{leavitt_1777_1908}. Over the past 20 years, time-domain observations from dedicated space-based telescopes have led to a revolution in astrophysics. Uninterrupted, high-precision light curves for millions of sources have affected fields ranging from solar system science, stellar astrophysics, exoplanet science, galactic stellar populations, supernovae, and galaxy formation. Astronomy has now entered an era in which a substantial number of stars in our galaxy have continuous high-precision space-based light curves, allowing large-scale variability classification efforts akin to stellar spectral typing in the early 20th century.  

This review will describe breakthroughs in astrophysics enabled by space-based time-domain telescopes. Particular focus will be given to results from NASA's Kepler/K2 \citep{borucki_kepler_2010,howell_k2_2014}  and TESS missions \citep{ricker_transiting_2015}, as opposed to time-domain observations from general purpose observatories such as the Hubble space telescope or Gaia. The review is intentionally broad in scope, covering high-level results across a range of topics with the goal to provide sufficient information on the current state of each field, but not replacing in-depth reviews. The review will only briefly cover exoplanet science and asteroseismology, which are the traditional scientific drivers for dedicated space-based time-domain telescopes, and have been the subject of other reviews \citep[e.g.][]{aerts_probing_2021,kurtz_asteroseismology_2022, lissauer_exoplanet_2023}.

\subsection{A Brief History of Space-Based Time-Domain Photometry} 

\noindent
Time-domain observations from space have long been recognized as a major opportunity to advance our understanding of exoplanets and stellar astrophysics. The first concepts for a dedicated time-domain monitoring of stars from space were put forward in Europe in the 1990's, including PRISMA \citep{appourchaux_prisma_1993}, STARS \citep{fridlund_stars_1995}, MONS \citep{kjeldsen_mons_2000} and Eddington \citep{favata_eddington_2004}. Eddington was an ESA-led mission with a planned 1.2\,m aperture to detect exoplanets using the transit method and perform asteroseismology of Sun-like stars. 
While it was selected, none of the above missions were ultimately funded for launch.

The first successfully launched mission dedicated to space-based time-domain photometry was the Canadian space telescope MOST \citep[Microvariability and Oscillations in Stars,][]{walker_most_2003,matthews_one_2007}. MOST was a 15\,cm telescope in a sun-synchronous low-Earth orbit, targeting individual stars for up to 90 continuous days with cadences as short as one minute with the aim to probe the structure of stars using asteroseismology. MOST successfully discovered pulsations in a range of intermediate-mass and massive stars, as well as oscillations in red giants \citep{kallinger_nonradial_2008}. MOST also measured rotational modulation in young stars \citep{croll_differential_2006} and detected some of the first exoplanet transits observed from space \citep{rowe_upper_2006, winn_super-earth_2011}.

The start of the space-photometry revolution is generally associated with the launch of the French-led  CoRoT (Convection Rotation and Planetary Transits) satellite in 2006 \citep{baglin_corot_2006}. CoRoT was also launched in a sun-synchronous orbit and carried out the first space-based ``long-stare'' campaigns lasting for nearly five months in two dedicated fields. CoRoT furthermore made several important advancements in exoplanet science, including the detection of a transiting rocky exoplanet \citep{leger_transiting_2009}. CoRoT also detected oscillations in several main-sequence stars \citep[see, e.g.,][]{michel_corot_2008} and in several thousands of red-giant stars \citep[see, e.g.,][]{de_ridder_discovery_2006,hekker_characteristics_2009}. 

\begin{figure}[t]
\begin{center}
\includegraphics[width=6in]{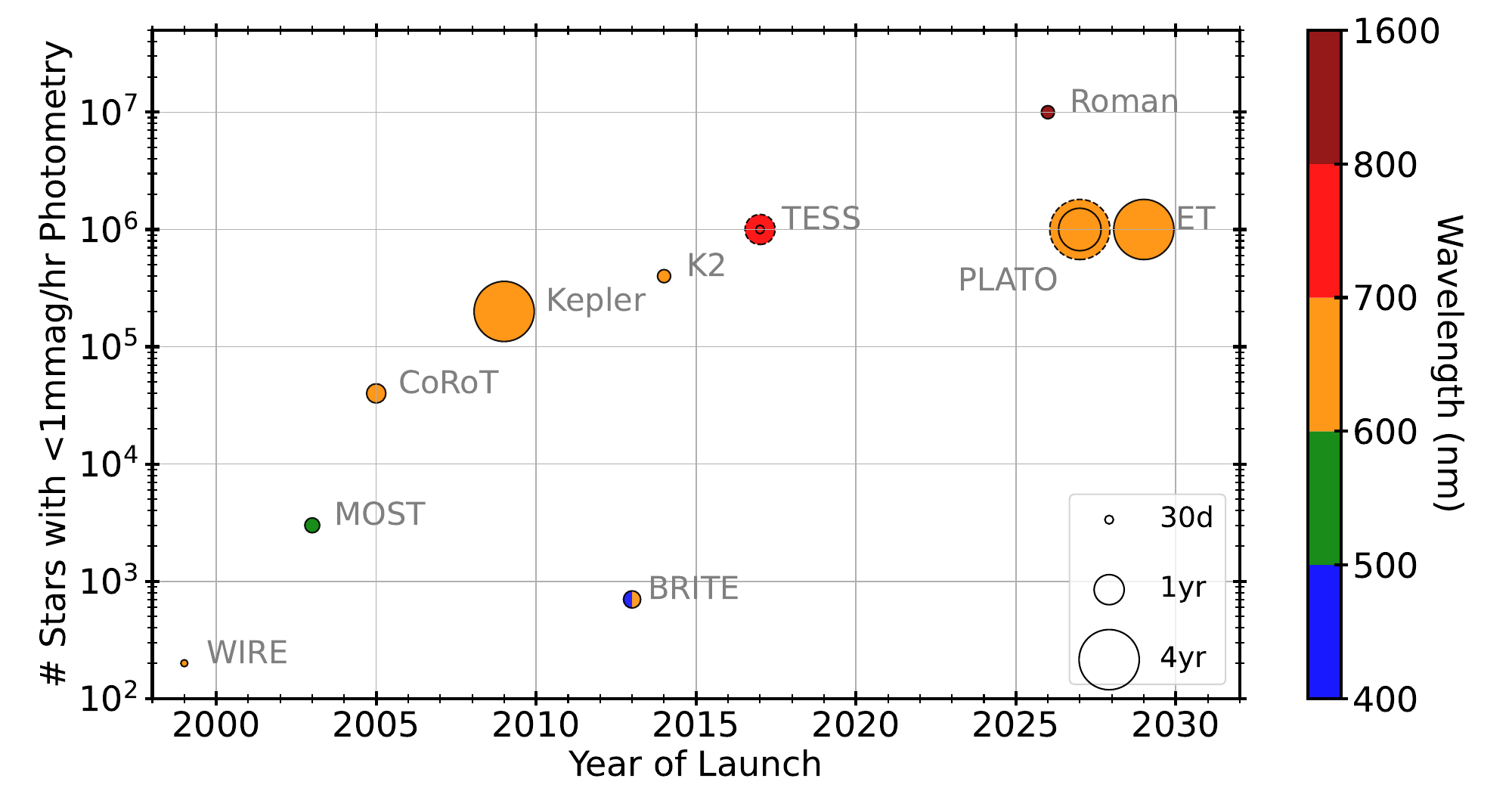}
\caption{\textbf{The number of stars with precise, continuous space-based light curves has increased by four orders of magnitude in the past 20 years.} Approximate number of stars observed with precisions of $<$1\,mmag per hour as a function of year of launch of space-based missions. Symbol areas scale with the typical time baseline of continuous observations. Dashed circles for TESS and PLATO denote the typical longest continuous observations. Color coding shows the wavelength of the central bandpass. BRITE Constellation is the only mission which observed with multiple bandpasses (CoRoT used a prism to obtain spectral information for exoplanet targets). See Table \ref{tab:space_telescopes} for details on yield estimates.}
\label{fig:history}
\end{center}
\end{figure}

In parallel, other space telescopes were used for time-domain astronomy. For example, the Hubble space telescope was used for both stellar variability studies \citep{edmonds_k_1996,gilliland_photometric_2008} and the characterization of transiting exoplanets \citep{nutzman_design_2008}. Furthermore, innovative approaches used ancillary data for stellar astrophysics, including the star-tracker of the failed WIRE  (Wide-Field Infrared Explorer) satellite \citep{schou_observations_2001,retter_oscillations_2003,bruntt_evidence_2005,stello_oscillating_2008}, as well as the SMEI (Solar Mass Ejection Imager) experiment \citep{tarrant_asteroseismology_2007}.
 
The launch of the Kepler space telescope in 2009 continued the revolution of space-based photometry. With the advantage of a large aperture, an Earth-trailing orbit, and continuous observations of a single field, Kepler provided light curves with unprecedented precision and duration for over 150,000 pre-selected stars \citep{borucki_kepler_2010, batalha_selection_2010}. Following the failure of two reaction wheels, Kepler ceased observations of its original field after nearly 4 years of operations. Thanks to engineering ingenuity, its successor mission K2 performed 70 day campaigns along the ecliptic plane, observing a total of nearly 400,000 pre-selected stars over the next 3 years \citep{howell_k2_2014}. The end of the Kepler/K2 era came in 2018, when spacecraft operations were discontinued due to lack of fuel, which prevented it from maintaining precise pointing control. 

BRITE constellation, a collaboration between Austria, Canada, and Poland, launched in 2013 to perform the first dedicated multi-color time-domain photometry from space \citep{weiss_brite-constellation_2014}. The constellation consists of six nanosatellites with 3\,cm apertures, each equipped either a red or blue filter, to perform precise two-colour photometry of naked-eye ($V<6$) stars. Key results so far include the study of mass loss in Be stars \citep{baade_short-term_2016}, the discovery of the most massive star in a heartbeat binary system to date \citep{pablo_most_2017}, and the detection of oscillations in nearby red giants \citep{kallinger_stellar_2019}.

The next phase of the space-photometry revolution was enabled by TESS, which launched in a lunar resonance orbit in 2018 \citep{ricker_transiting_2015, winn_transiting_2024}. TESS is largely conducting an all-sky survey strategy, obtaining on average 27-day light curves with continuous coverage for up to one year near the ecliptic poles. The proximity to Earth allows TESS the collection of full-frame images, thus removing the need for target selection except for stars which require the fastest cadences. At the time of writing, TESS is monitoring 2000 stars per observing sector in 20-second cadence, tens of thousands of stars in 2-min cadence, and obtains full frame images at 200-second cadence\footnote{Full frame images were obtained at 30-minute cadence during the first 2 years of the TESS Mission, and at 10-minute cadence during years 3-4.}.

Figure \ref{fig:history} summarizes the above ``brief history'' of space-based photometry by comparing the number of stars with $<$1mmag/hr photometric precision as a function of launch date of the mission\footnote{The figure excludes CHEOPS \citep{benz_cheops_2021}, which is a targeted follow-up mission.}. Over the past 20 years, the number of stars with such data has increased by 4 orders of magnitude. The next chapters of the space-photometry revolution will be written by ESA's PLATO \citep{rauer_plato_2025}, NASA's Roman \citep{spergel_wfirst-24_2013}, and the Chinese Earth 2.0 \citep{ge_et_2022} missions, which are slated for launch in the second half of this decade. Chapter \ref{ch:future} provides a brief outlook on these new time-domain missions.

\subsection{Fundamental Parameters of Space-Based Time-Domain Missions}

\noindent
Space-based time-domain missions can be characterized by a set of fundamental parameters, which set the framework for achievable science with a given mission. This section gives a pedagogical introduction into the most common parameters of time-domain missions.

\begin{itemize}

\item The \textit{photometric precision} is the relative uncertainty of brightness measurements. The detected signal from a star is given by:

\begin{equation}
S = F_{*} \times t \times \pi (D/2)^{2} \times (h\nu)^{-1} \times Q \: .
\end{equation}

Here, $S$ is the total number of electrons or counts received from the star, $F_{*}$ is the flux emitted per unit time and unit area, $t$ is the integration time, $D$ is the telescope diameter, $h\nu$ is the photon energy and $Q$ includes instrument characteristics such as detector quantum efficiency and system throughput. In the idealized limit of photon noise, the measurement error or noise is $N=\sqrt{S}$. The photometric precision, which is the relative error on a flux measurement, is thus given by $\sigma = N/S = S^{-1/2}$. For a fixed bandpass and detector we therefore have:

\begin{marginnote}[]
\entry{Photometric Precision}{Fractional uncertainty of brightness measurements; in the limit of photon noise, the precision increases with telescope aperture, integration time, and stellar brightness.}
\end{marginnote}


\begin{equation}
\sigma \propto F_{*}^{-1/2} \times t^{-1/2} \times D^{-1} \: .
\label{ch1:equ:rms}
\end{equation}

As expected, better photometric precision (lower $\sigma$) is achieved for larger telescope apertures, longer integration times and brighter stars. Equation \ref{ch1:equ:rms} allows simple scalings from one mission to another. For example, Kepler's aperture is approximately ten times larger than TESS's. Therefore, for a fixed brightness of a star, the photometric precision per unit time (fixed cadence) is 10 times better for Kepler compared to TESS. Equivalently, a star needs to be 100 times (5 magnitudes) brighter for TESS to reach the same photometric precision per unit time as Kepler. Equation \ref{ch1:equ:rms} is a good approximation for space-based photometry, which does not suffer from scintillation noise, but breaks down for bright stars due to saturation and at the faint stars due to sky noise and detector read noise. 

Space-based time-domain missions typically collect data in sub-exposures that are co-added on board to a given cadence, which is then downloaded to the ground. For example, for Kepler the exposure time for individual frames was 6 seconds, while for TESS individual exposure times are 2 seconds. This explains why the light curve scatter for a given star increases as the time sampling of the data product increases (Figure \ref{fig:scatter}). For example, comparing the light curve of the same star observed by TESS with 30-minute cadence and 20-second time sampling shows that the light curve scatter of the former is smaller by a factor $\sqrt{30\times60/20} \approx 10$, since each data point is based on a longer effective exposure time. In other words, the light curve scatter per unit time remains constant for a given instrument.

\begin{figure}[t]
\begin{center}
\includegraphics[width=6in]{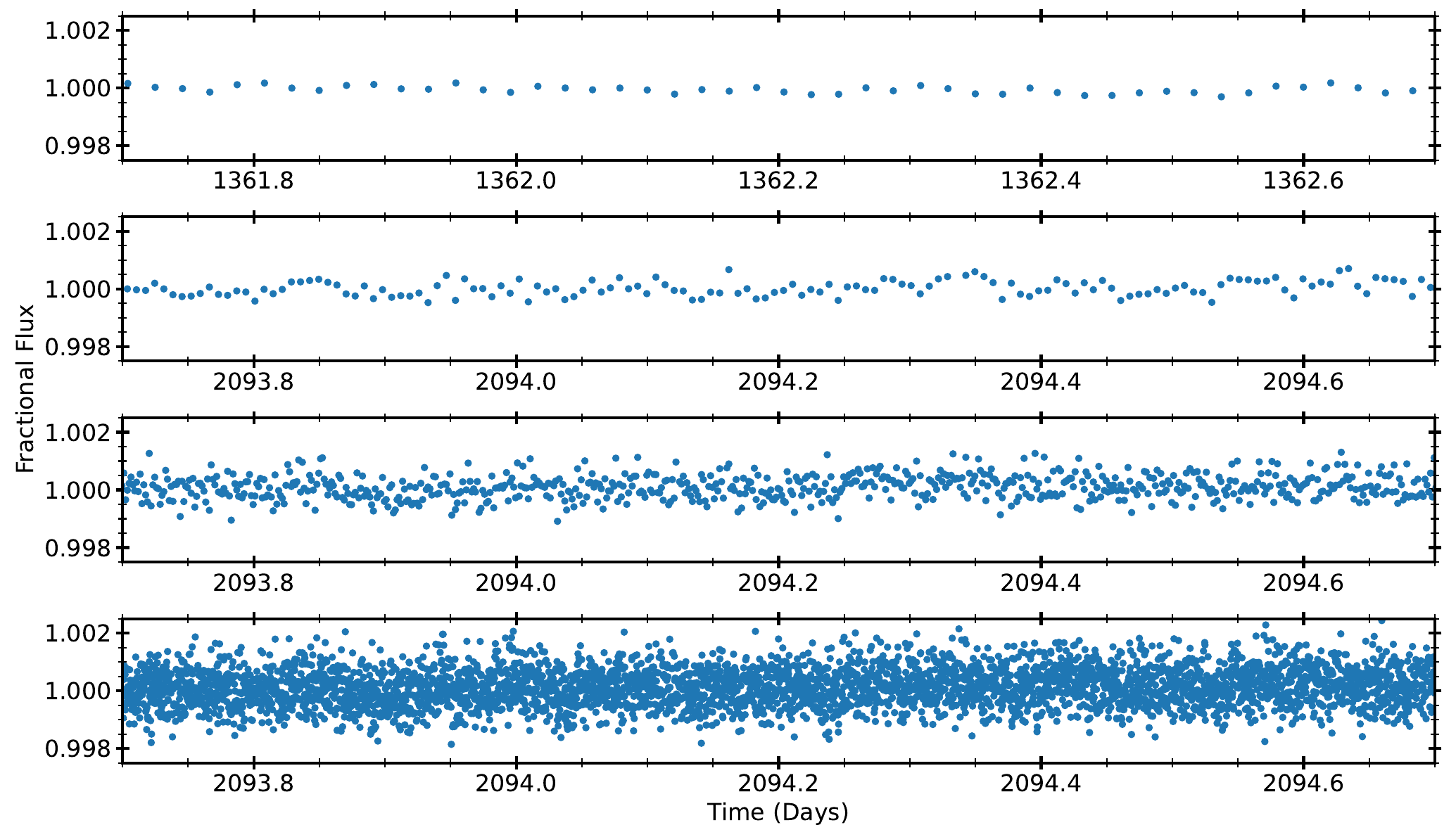}
\caption{\textbf{For a given space-based time-domain telescope, the light curve scatter per unit time is constant.} Each panel shows a 12 hour segment of the same star using a different TESS data product: 30-min sampling (top), 10-minute sampling (second from top), 2-min sampling (second from bottom) and 20-second sampling (bottom). The x-axis and y-axis scales are identical in all panels. }
\label{fig:scatter}
\end{center}
\end{figure}

\item The \textit{time baseline} is the typical length of a light curve. It sets a limit on the lowest frequency (longest period) that can be unambiguously measured from a light curve. For evenly spaced data, the frequency resolution in the Fourier domain is given by:

\begin{equation}
\Delta f = 1/T \: .
\label{ch1:equ:fres}
\end{equation}

In frequency analysis, Fourier spectra are commonly oversampled by a factor proportional to $\Delta f$ (typically $5-10$).

\begin{marginnote}[]
\entry{Time Baseline}{Length of uninterrupted observations; sets lowest measurable frequency (longest detectable period).}
\end{marginnote}

The maximum time baseline is determined by the spacecraft orbit. Continuous observations of a single field in a sun-synchronous orbit around Earth (such as for MOST, CoRoT, BRITE) are typically limited to $\approx$\,5 months due to Earth's orbital motion around the Sun. Kepler, located in an Earth-trailing orbit, allowed uninterrupted observations for fields off the ecliptic plane, with quarterly rolls to keep the spacecraft oriented towards the Sun. The innovative lunar resonance orbit of TESS allows uninterrupted observations near the ecliptic poles, while engineering requirements related to illumination of solar panels and thermal stability restrict typical observing lengths to $\approx$27 days. Planned missions such as Roman, PLATO and Earth 2.0 will be deployed in an orbit around the second Lagrange point (L2), allowing long uninterrupted observations over large parts of the sky. Disadvantages of this solution are higher mission costs, driven by the requirement for larger antennas to download data back to Earth, and overall lower data rates.

The time baseline can refer to the length of continuous observations or the total length of timeseries. The difference is the duty cycle, which is the fraction of time observations have been taken between the first and the last datapoint. For example, Kepler has collected the (so-far) longest continuous timeseries to date (4 years with a duty cycle close to 90\%), while TESS is currently obtaining the longest total timeseries ($>$ 8 years with a maximum duty cycle of $\approx$\,50\%).

\item The \textit{observing cadence} or \textit{time sampling} is the number of observations per unit time that are downloaded from the spacecraft\footnote{``Cadence'' can sometimes be used in a confusing manner. For example, light curves are frequently referred to as having ``30-minute cadence'' or ``1-minute cadence'', but the latter has a higher cadence than the former. Alternative wording such as ``sampling'' or ``faster/slower cadence'' can avoid this ambiguity.}. The cadence sets a fundamental limit on the highest frequency (or shortest period) of signal that can be recovered from a light curve. The upper limit for a time sampling $\Delta t$ is given by the Nyquist frequency:

\begin{equation}
f_{\rm{Nyq}} = 1/ 2 \Delta t \; .
\end{equation}

The Nyquist frequency is only strictly defined for equally sampled timeseries, and signals above the Nyquist frequency can be recovered from typical Kepler/K2 and TESS light curves \citep{murphy_super-nyquist_2013, chaplin_super-nyquist_2014, liagre_beyond_2025}.

\begin{marginnote}[]
\entry{Cadence or Sampling}{Number of observations per unit time; determines highest detectable frequency (Nyquist limit).}
\entry{Nyquist Frequency}{Highest frequency reliably detectable given the sampling cadence; signals above are aliased.}
\entry{Field of View (FOV)}{Angular sky area observed by a telescope; larger FOV enables more targets at lower spatial resolution.}
\entry{Pixel Scale}{Angular size covered by one detector pixel; balances field of view with spatial resolution.}
\end{marginnote}

The time sampling also causes fractional amplitude attenuation caused by time-averaging of a signal with frequency $f$, which is given by:

\begin{equation}
A = {\rm sinc}(\pi f t_{\rm exp}) \; ,
\end{equation}

where ${\rm sinc}(x) = (\sin x)/x$ and $t_{\rm exp}$ is the exposure time. For observations with no dead time between exposures (such as those obtained by Kepler/K2 and TESS), the exposure time is equal to the sampling time and thus the fractional attenuation in power can be written as:

\begin{equation}
P = {\rm sinc}^2\left({\pi \over 2} {f \over f_{\rm nyq}} \right) \; ,
\label{equ:poweratt}
\end{equation}

In the time-domain sampling can cause smearing effects, for example for the ingress and egress of transit or eclipse observations.

\item The \textit{number of targets} specifies the number of light curves produced per observing interval for a given cadence. This is predominantly set by the onboard storage capacity of the spacecraft, as well as the capability to download data to the ground and thus the orbit and communications antenna. Missions typically download smaller number of targets with faster sampling, and a larger number of targets at slower sampling.

\item The \textit{field of view and pixel size} set the spatial scales over which stars can be observed and resolved. The plate scale of an instrument is given by:

\begin{equation}
s[''/mm] = \frac{206.265}{F[mm]} = \frac{206.265}{f D[mm]} \; ,
\label{equ:platescale}
\end{equation}

where $F$ is the focal length, $D$ is the telescope diameter, $f=F/D$ is the f-ratio. For a given physical size and pixel dimensions of a detector, Equation \ref{equ:platescale} allows the calculation of the pixel size and field of view. For example, a single TESS detector has 4096$\times$4096 pixels with a physical size of 62$\times$62\,mm. The focal length of $F=146$\,mm for a single TESS camera yields a pixel size of $62/4096\times206.265/146\approx21$'' and field of view of $62/3600\times206.265/146\approx24$ degrees \footnote{The calculation does not take into account optical design considerations required to ensure aberrations remain small over a large FOV.}.

Blending of stars within a given aperture causes the dilution of a  signal as follows:

\begin{equation}
A_{d} = \frac{A}{(1+10^{-0.4 \Delta m})} \: ,
\end{equation}

where $A$ is the original signal amplitude, $A_{d}$ is the diluted amplitude and $\Delta m$ is the magnitude difference between the two stars in a given observing bandpass.

\begin{marginnote}[]
\entry{Bandpass}{Wavelength range of observations; affects photometric precision and stellar types most effectively studied.}
\entry{Blending (Dilution)}{Flux contamination from nearby stars within an aperture, reducing observed amplitudes.}
\end{marginnote}

\item The \textit{observing bandpass} determines wavelength range over which photons are collected. The majority of space-based time-domain missions used broadband optical bandpasses in order to collect as many photons as possible, resulting in higher photometric precision (see Equation \ref{ch1:equ:rms}). TESS used a redder bandpass than Kepler/K2 to improve sensitivity for cool main-sequence stars. Multiple photometric bands can aid in a number of astrophysical contexts such as asteroseismic mode identification and temperature measurements of spots. The BRITE nanosatellite constellation is so far the only space-based time-domain mission to have used dedicated multiple filters. Some mission concepts that will make use of multiband photometry are currently under consideration (see Section 7).

\end{itemize}

In practice, many of the above parameters are linked. For example, there is a fundamental ``deep and narrow'' versus ``shallow and wide'' tradeoff between aperture size (photometric precision) and field of view, and therefore number of observable targets. Kepler's larger aperture resulted in a smaller FOV allowed with a smaller pixel size (4''), whereas the large field of view from TESS comes at the cost of 21'' pixels and a 10 times smaller aperture. On the other hand, the proximity of TESS orbit to Earth allows data to be downloaded at a higher rate for a given antenna size, while the number of stars observed by Kepler was limited to data downlink rate from an Earth-trailing orbit. Table \ref{tab:space_telescopes} summarizes the parameters for missions shown in Figure \ref{fig:history}.

\begin{table}[h]
\tabcolsep7.5pt
\caption{Observational parameters of past, current and approved space-based time-domain missions.}
\tabcolsep7.5pt
\label{tab:space_telescopes}
\begin{center}
    \begin{tabular}{lccccccccc}
        \hline
        Mission & Launch & T$_{\rm max}$ & D & FOV & Res & $\Delta$t & T$_{\rm cont}$ & $\lambda$ & n \\
        --- & --- & yrs & cm & (deg$^2$) & ''/pix & sec & days & nm & --- \\
        \hline
        WIRE    & 1999 & 6  & 5    & 64    & 60 & 15 & 20 & 620  & $\sim$ 200 \\
        MOST    & 2003 & 15.8 & 15   & 5   & 3   & 30  & 60 & 550   & $\sim$ 3000 \\
        CoRoT   & 2006 & 7.4  & 27   & 7.6    & 2.3 & 32, 512  & 150 & 650   & $\sim 4\times 10^{4}$ \\
        Kepler  & 2009 & 4.0 & 95   & 105    & 4 & 60, 1800 & 1460 & 600   & $\sim 2\times 10^{5}$ \\
        BRITE   & 2013 & $>$12.6$^{\dagger}$ & 3    & 24     & 28  & 10  & 180   & 420, 620 & $\sim$ 700 \\
        K2      & 2014 & 4.4 & 95   & 105    & 4 & 60, 1800 & 70  & 600   & $\sim 5\times 10^{5}$ \\
        TESS    & 2018 & $>$7.5$^{\dagger}$ & 4$\times$10   & 2300   & 21  & 20, 120, 600 & 27 & 800   & $\sim 10^{6}$ \\
        \hline
        Roman   & 2026 & 5$^{\ddagger}$ & 240  & 0.28   & 0.11 & 720 & 70 & 1290  & $\sim 10^{7}$ \\
        PLATO   & 2027 & 4$^{\ddagger}$ & 12$\times$26 & 2250   & 15  & 2.5, 25, 50, 600 & 1460 & 750 & $\sim 10^{6}$ \\
        Earth2.0& 2028 & 4$^{\ddagger}$ & 6$\times$30 & 500   & 4  & 30, 600, 1800 & 1460 & 675 & $\sim 10^{6}$ \\
        \hline
    \end{tabular}
\end{center}
\begin{tabnote}
Launch dates for Roman, PLATO and Earth 2.0 are estimated. K2 was a continuation of the Kepler mission, and thus the year refers to the start of operations rather than a launch date. $T_{\rm max}$ denotes the mission duration, and thus the upper limit on the longest possible time baseline of observations ($^{\dagger}$current missions that are still active, $^{\ddagger}$nominal durations for planned missions excluding possible extensions). FOV is the total field of view. $\Delta t$ and T$_{\rm cont}$ are the typical cadence and duration of continuous light curves, $n$ is the number of observed stars with $<$1mmag/hr photometry, $\lambda$ is the central wavelength of the observing bandpass. For Roman, FOV and T$_{\rm cont}$ are taken from the Galactic Bulge Time Domain Survey. For missions up to and including K2 $n$ is assumed to be approximately the total number of observed stars (since the magnitude corresponding to a $<$1mmag/hr photon noise limit is fainter than most targeted stars). For TESS, $n$ was estimated as the total number of stars in the TESS Input Catalog down to $I<12$\,mag. For Roman, $n$ was estimated as the total number of star counts with $H<18$\,mag based on the photometric precision model from \citet{penny_predictions_2019} and the synthetic population from \citet{gould_wfirst_2015}. For PLATO and Earth 2.0, $n$ was taken as the estimated number of stars with downloaded light curves from \citet{rauer_plato_2025} and \citet{ge_et_2022}.
\end{tabnote}
\end{table}

Systematic errors are another important characteristic of space-based missions. For time-domain missions, important examples include:

\begin{itemize}
    \item Gaps due to periodic downlinks of data to Earth, which can introduce aliasing and affect the recovery of periodicities in the data.
    \item Spacecraft motion, which causes periodic changes in the point-spread function, aperture flux losses and intrapixel sensitivity changes. This was a challenge for K2, which only operated with two reaction wheels and thus experienced periodic drifts.
    \item Pointing inaccuracies, which cause changes in the position of stars on the detector after each observing segment. This can lead to offsets in absolute flux, requiring ``stitching'' of data segments, and causing difficulty in measuring small amplitude signals with periods longer than an observing quarter/sector.
    \item Background variations due zodiacal dust and straylight reflected off the Earth or the Moon. This causes periodic flux changes which introduce systematic errors.
\end{itemize} 

Some systematic errors are influenced by mission parameters. For example, spacecraft orbits close to Earth increase the effects of background variations due to straylight entering the telescope aperture. Many of the systematics mentioned above can be successfully mitigated with sophisticated data reduction techniques \citep[e.g.][]{reegen_reduction_2006}.

\subsection{From Exoplanets to Astrophysics: the Evolution of Target Selection} 

\begin{figure}[t]
\begin{center}
\includegraphics[width=6in]{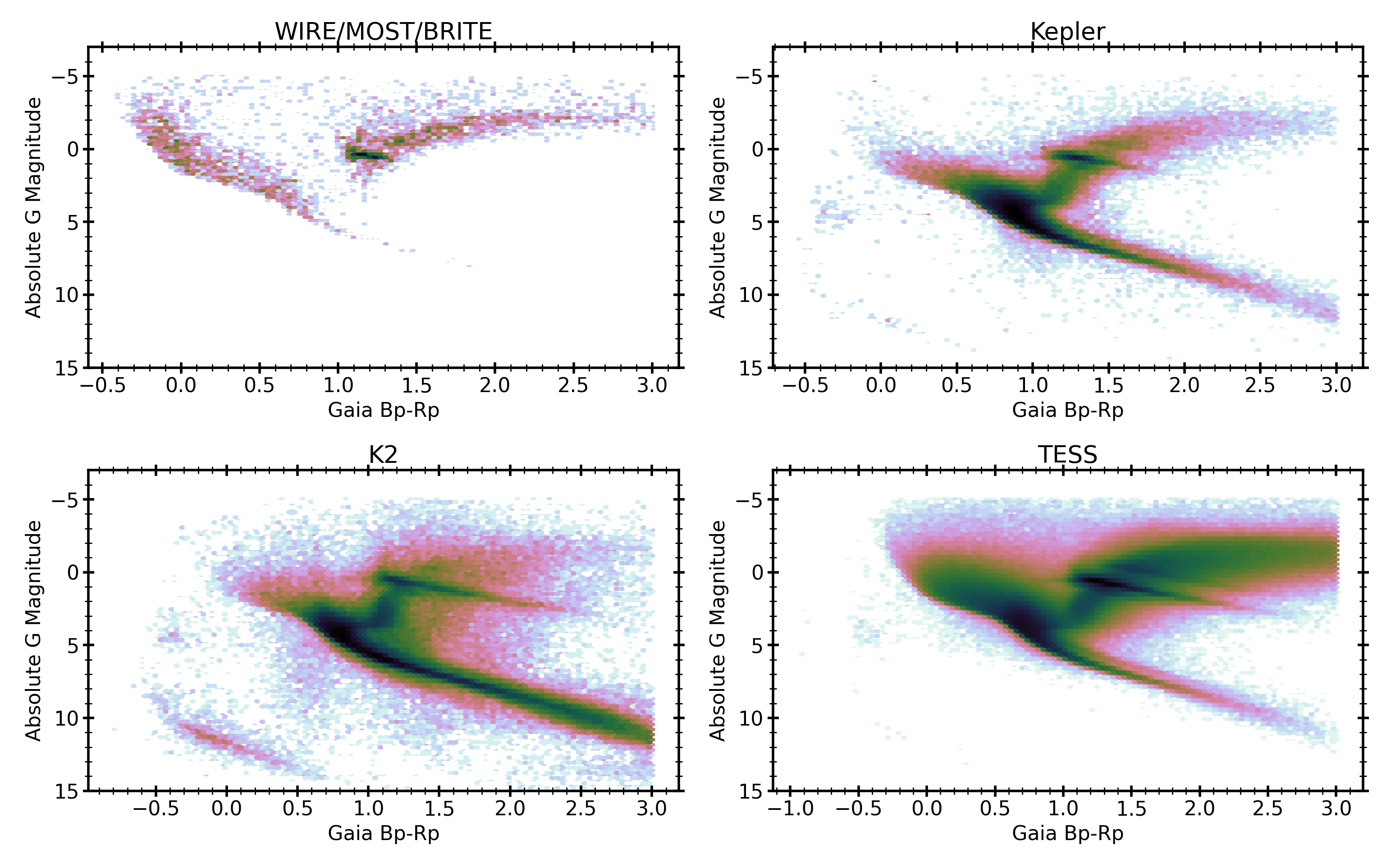}
\caption{\textbf{The evolution of target selection for space-based photometry missions.} Gaia color versus absolute magnitude for stars observed by space-based time-domain mission. Color-coding shows the logarithmic number density in each bin. WIRE, MOST and BRITE distributions are approximated by stars with $G < 6$, and TESS distribution with $G < 12$. The Kepler and K2 panels use actual target lists. Values are not corrected for reddening.}
\label{fig:cmd}
\end{center}
\end{figure}

\noindent
A major component for space-based time-domain mission is target selection, since onboard storage capacities are frequently insufficient to download all the data that is collected. Figure \ref{fig:cmd} illustrates how target selection differs between missions. Missions such as WIRE, MOST and BRITE were predominantly magnitude limited, resulting in a preference for luminous, massive main-sequence or giant stars. Kepler targeted main-sequence solar-like stars to detect Earth-like planets \citep{batalha_selection_2010, wolniewicz_stars_2021}, but also observed many red-giant stars because of the difficulty of discerning dwarf from giant stars in the absence of parallaxes in the pre-Gaia era \citep{mann_they_2012}. While these stars yielded a wealth of stellar astrophysics results, they also contended with exoplanet science: many giant stars were dropped from the target list based on their variability characteristics in favor of smaller stars that are more suitable for the detection of exoplanets. 

The rich astrophysics science yield from Kepler shaped the target selection for later missions. The K2 Mission targeted a broader distribution of spectral types \citep{huber_k2_2016}, including compact stars and massive stars, driven by the exoplanet-science focus shifting to cooler dwarfs and more diverse general astrophysics investigations. TESS full-frame images have mostly removed the need for target selection. An exception are stars requiring very fast cadence such as white dwarfs, flare stars and oscillating Sun-like stars.

\section{Stellar Astrophysics}

\subsection{Asteroseismology} 

\noindent
Asteroseismology is the study of pulsations to probe the interior of stars. Stellar pulsations are observed across the Hertzsprung-Russell (HR) diagram and can be broadly divided into driving mechanisms due to opacity changes in the stellar interior (so-called classical pulsators such as $\delta$\,Scuti stars and Cepheids) and stars which oscillate due to turbulent surface convection (so-called solar-like oscillators). Stellar pulsations span a wide range of timescales (minutes to years) and amplitudes (few parts-per-million to several percent), and as such are exquisitely matched to the continuous, high-precision light curves produced from space-based telescopes. A number of reviews have been dedicated to the scientific advances in asteroseismology over the past decade, including solar-like oscillators \citep{chaplin_asteroseismology_2013, garcia_asteroseismology_2019, jackiewicz_solar-like_2021} and classical pulsators \citep{guzik_highlights_2021, kurtz_asteroseismology_2022}. This section provides a brief historical overview on the impact of the space-based photometry revolution on asteroseismology, with a focus on solar-like oscillators.

\begin{marginnote}[]
\entry{Asteroseismology}{Study of stellar oscillations to probe internal structure, ages, and fundamental parameters.}
\end{marginnote}

Prior to space-based photometry, the detection of solar-like oscillations relied on ground-based radial velocities \citep[see ][for a review]{bedding_solar-like_2003}. The first confirmed detection of oscillations in a star other than the Sun was made in Procyon by \citet{brown_detection_1991}, followed by the first  detection of regularly spaced frequencies in $\eta$\,Boo by \citet{kjeldsen_solarlike_1995}. The greatly improved radial velocity precision for detecting exoplanets enabled the detection of oscillations in several nearby main sequence and subgiant stars such as $\beta$\,Hyi \citep{bedding_evidence_2001,carrier_solar-like_2001}, $\alpha$\,Cen\,A \citep{bouchy_p-mode_2001, butler_ultra-high-precision_2004} and B \citep{carrier_solar-like_2003, kjeldsen_solar-like_2005}, as well as red giant stars such as $\xi$\,Hya \citep{frandsen_detection_2002} and $\epsilon$\,Oph \citep{de_ridder_discovery_2006}. 


However, the short timeseries and gaps in ground-based observing campaigns limited their potential. Space-based photometry from MOST initially yielded a non-detection in Procyon \citep{matthews_no_2004}, with later datasets showing a detection consistent with radial velocity observations \citep{guenther_nature_2008, huber_solar-like_2011}. MOST also detected oscillations in red giants \citep{barban_detection_2007}, including the first observational evidence for non-radial modes \citep{kallinger_nonradial_2008}. CoRoT achieved a major breakthrough by detecting oscillations in a number of main sequence stars \citep[e.g.][]{appourchaux_corot_2008, michel_corot_2008} and several thousands of red giant stars 
\citep[e.g.][]{hekker_characteristics_2009}. 
Importantly, CoRoT unambiguously demonstrated that red giants  oscillate in non-radial modes \citep{de_ridder_non-radial_2009}, which opened the door for detailed  studies of the interior structure of red giants \citep[see][for a recent review]{hekker_giant_2017}.

\begin{figure}
\begin{center}
\includegraphics[width=6in]{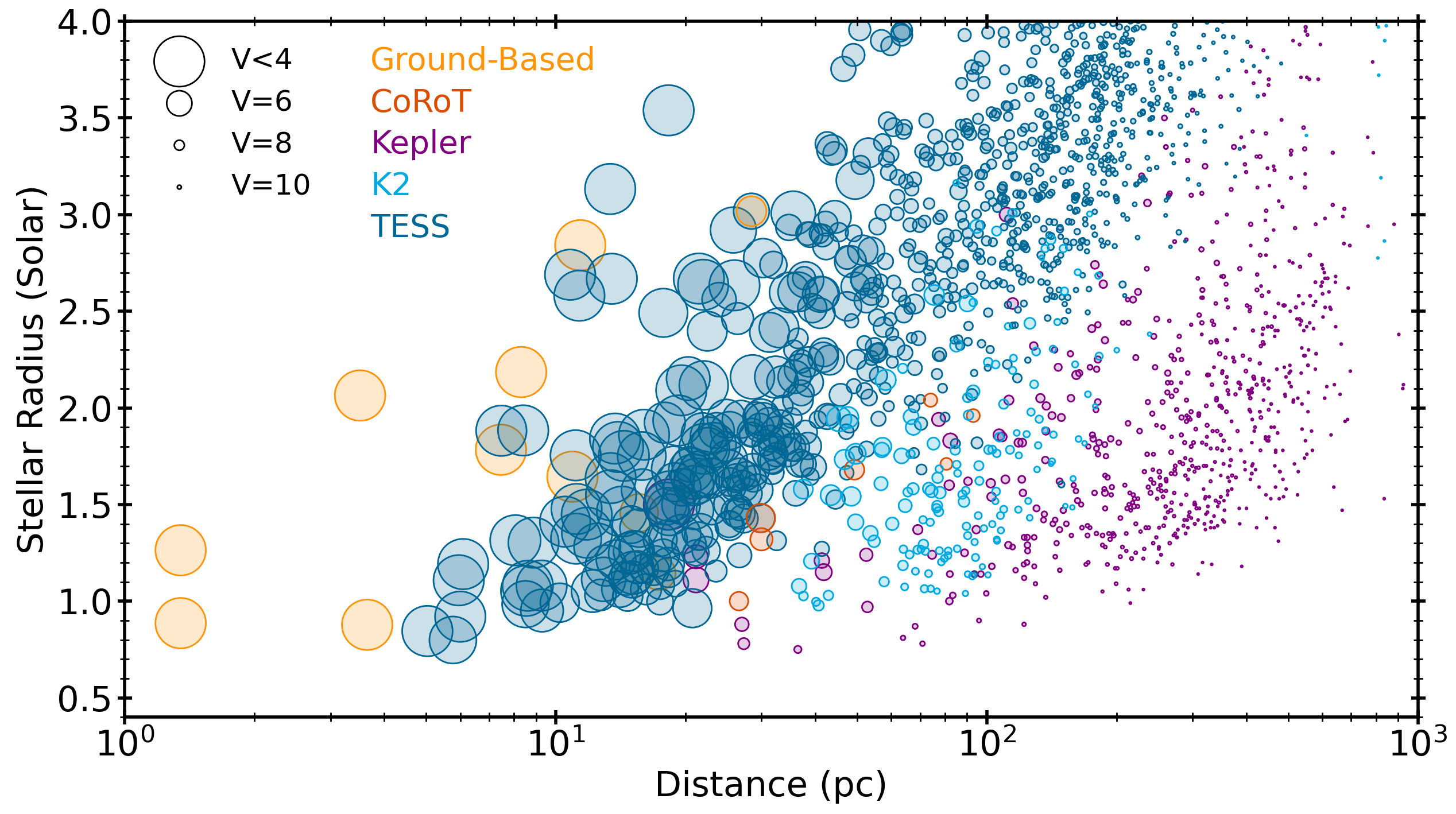}
\caption{\textbf{Stars with detected solar-like oscillations as a function of distance.} Following ground-based RV campaigns ($\approx$\,20 stars), CoRoT yielded the first high-quality detections from space ($\approx$\,10 stars). Kepler enabled $\approx$\,700 detections, but mostly in faint and distant stars. The K2 ($\approx$\,200 detections) and TESS ($\approx$\,900 detections) missions have bridged the gap to nearby stars, for which asteroseismology can be combined with complementary methods. Catalog detections are taken from \citet{sayeed_homogeneous_2025} (Kepler), \citet{lund_k2_2024} (K2), \citet{hatt_catalogue_2023} and \citet{lund_luminaries_2025} (TESS).}
\label{fig:seismo2}
\end{center}
\end{figure}

 \textit{Kepler} detected oscillations in over 600 main-sequence and subgiant stars \citep{chaplin_asteroseismic_2014, mathur_detections_2022, sayeed_homogeneous_2025} and over twenty thousand red giants \citep{hekker_characterization_2011, stello_asteroseismic_2013, mathur_probing_2016, yu_asteroseismology_2018}, enabling the study of oscillations across the low-mass H-R diagram. The larger number of red giants with detected oscillations is due to a combination of two effects: First, oscillation amplitudes increase with luminosity \citep{kjeldsen__bedding_amplitude_1995}, making a detection easier at a given apparent magnitude. Second, the majority of  targets were observed with 30-minute sampling, setting an upper limit of $\log g \sim 3.5$ since less evolved stars oscillate above the Nyquist frequency.

TESS has continued the solar-like oscillator revolution in two important ways. First, TESS observes main-sequence and subgiant stars that are much brighter and nearby than those detected by Kepler (Figure \ref{fig:seismo2}). This allows the combination of asteroseismology with classical observations such as long-baseline interferometry, which can constrain important physics such as the convective mixing length parameter \citep[e.g.][]{silva_aguirre_standing_2017} and initial Helium abundances \citep[e.g.][]{verma_helium_2019}.
TESS's 20-second cadence mode, introduced in the first extended mission, has led to an increase in photometric precision of bright stars \citep{huber_20_2022}, enabling the detection of oscillations in hundreds of unevolved stars \citep{hatt_catalogue_2023, lund_luminaries_2025}. Second, TESS has enabled the near all-sky detection of oscillating red giants, which are critical probes for galactic archeology (Section \ref{sec:ga}). As of writing, TESS has detected oscillations in 160,000 red giants \citep{hon_quick_2021}, an order of magnitude increase over Kepler, based on the analysis of only single sectors of data over most of the sky.

Space-based time-domain missions also revolutionized asteroseismology of classical pulsators such as $\delta$\,Scuti stars, $\gamma$\,Doradus stars, pulsating OB stars and compact pulsators. Examples include the first detection of regular gravity-mode period spacings in a massive star by CoRoT \citep{degroote_deviations_2010} and breakthroughs in the Kepler/K2 era such as the discovery of hybrid $\delta$\,Scuti stars - $\gamma$\,Doradus stars \citep[e.g.][]{grigahcene_hybrid_2010}, core rotation period measurements for hundreds of intermediate-mass stars \citep{van_reeth_interior_2016, li_gravity-mode_2020},  and measurements of dozens of rotation periods for white dwarfs \citep[e.g.][]{hermes_white_2017}. The TESS Mission expanded the revolution to a larger sample of classical pulsators, which were sparse in Kepler/K2, leading to the discovery of stochastic low-frequency variability in massive OB stars \citep{bowman_low-frequency_2019}, constraints on interior mixing mechanisms \citep{pedersen_internal_2021} and the discovery of regular spacings in young $\delta$\,Scuti stars \citep{bedding_very_2020}. For more extensive reviews of classical pulsators the reader is referred to \citet{kurtz_asteroseismology_2022} and \citet{bowman_asteroseismology_2024} for general reviews, \citet{aerts_asteroseismic_2024} for fast rotators and \citet{corsico_pulsating_2019} for pulsating white dwarfs.

\subsection{Convection and Granulation}

\noindent
Convection is one of two major energy transport mechanisms in stars. In stellar layers that are unstable to convection, an equilibrium perturbation will result in gas parcels continuously rising and thus transporting energy from the interior to the surface. The conditions for unstable convection depend on the temperature and composition gradients within the star \citep{schwarzschild_structure_1958, ledoux_stellar_1947}. Broadly speaking, the conditions for convective energy transport are found in the cores of stars with masses $\gtrsim 1.2 M_{\odot}$ and in the envelopes of stars with masses $ \lesssim 1.2 M_{\odot}$.

\begin{marginnote}[]
\entry{Granulation}{Surface brightness fluctuations caused by convection; linked to stellar surface gravity and oscillations.}
\end{marginnote}

Stellar granulation is a manifestation of convection on the surfaces of stars. As convective cells rise and cool, the integrated brightness of stars as observed in light curves change over time. Granulation causes brightness variations over a broad range of timescales. \citet{harvey_high-resolution_1985} postulated that the time evolution of granulation can be approximated by a series of Lorentzian functions with typical decay timescales, which correspond to granules with typical sizes. These timescales are commonly referred to as granulation (smallest scales), mesogranulation (larger scales), and supergranulation (largest scales). 

\begin{figure}
\begin{center}
\includegraphics[width=3in]{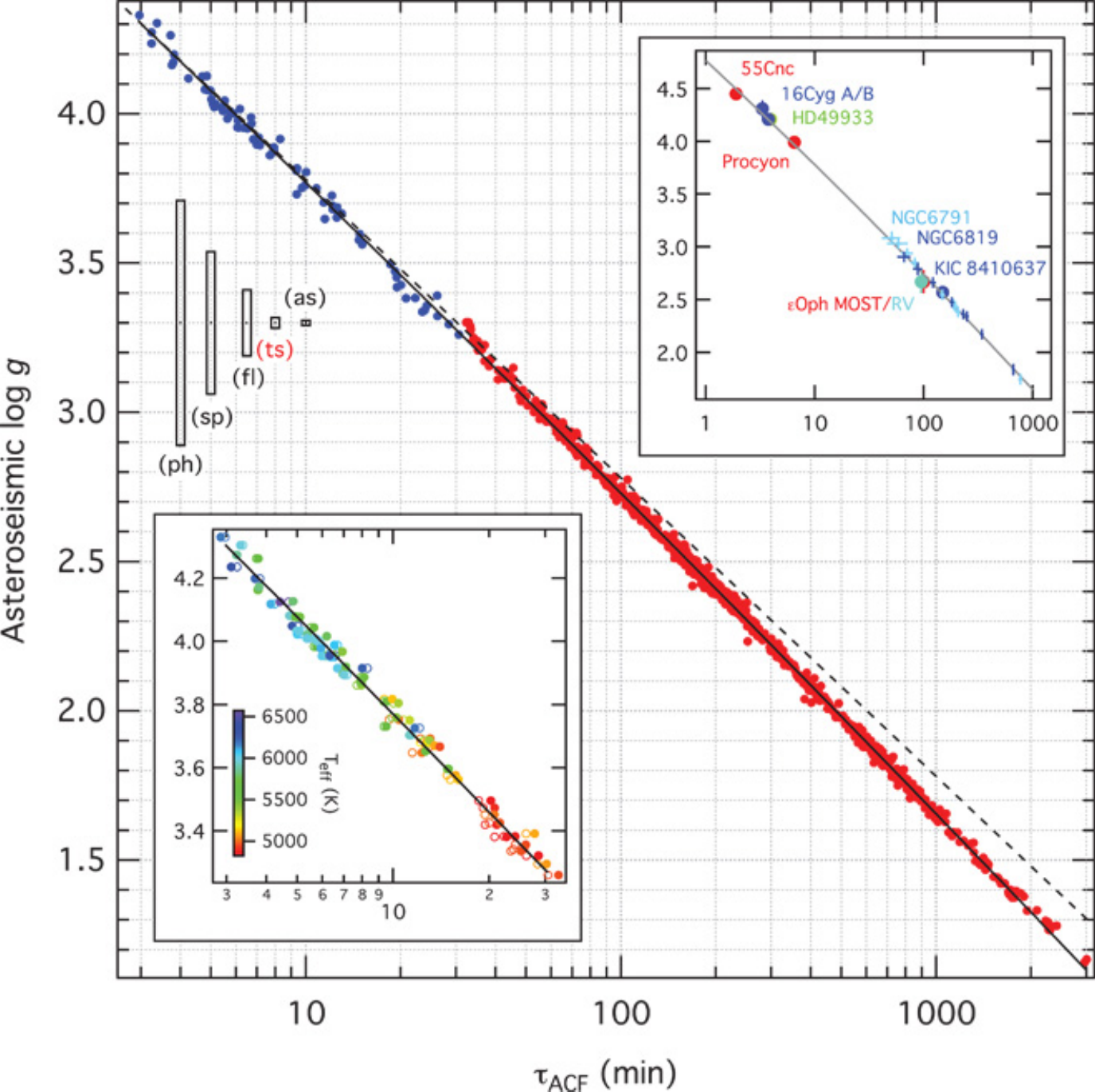}
\caption{\textbf{Space-based telescopes revealed that the time-domain variability due to stellar granulation and oscillations are closely linked.} Correlation between surface gravities measured through asteroseismology and the autocorrelation of light curves for stars observed by Kepler. Figure reproduced from \citet{kallinger_precise_2016}.}
\label{fig:gran}
\end{center}
\end{figure}

Informed by numerical simulations \citep{freytag_hydrodynamical_1996}, \citet{kjeldsen_amplitudes_2011} proposed scaling relations for the timescales and amplitudes of stellar granulation. The assumption is that both stellar oscillations and granulation are driven by convection, and thus the typical timescales and amplitudes scale linearly with the pressure scale height $H$, which for an isothermal atmosphere is given by $H \propto g^{-1}$, with $g$ denoting the surface gravity of the star. This is the same dependence as the typical timescale of stellar oscillations (the frequency of maximum power, $\nu_{\rm max}$). 

A major breakthrough of the space-based photometry revolution was the observational confirmation that granulation indeed shows a tight scaling with the properties of stellar oscillations, and thus fundamental parameters of stars. \citet{mathur_granulation_2011} first demonstrated that both the timescales and amplitudes of granulation in Kepler red giants show a tight correlation to the frequency of maximum power and thus stellar surface gravity. For main-sequence and subgiant stars, \citet{bastien_observational_2013} derived an empirical relation between the brightness variations at a fixed 8-hour timescale (so-called ``flicker''). Further investigation demonstrated that granulation tracks surface gravity as precisely as oscillations \citep{kallinger_connection_2014}, allowing the determination of surface gravities to a precision of $\approx$\,0.01 dex over an astounding 5 orders of magnitude in surface gravity (Figure \ref{fig:gran}).

The tight connection between granulation and oscillations opened up a flurry of new methods to precisely measure surface gravities from light curves alone.  This is significant because granulation extends over a wide range of timescales, and thus does not require the fast sampling that is required to measure oscillations. Furthermore, the combination of surface gravities with precise radii of stars derived from Gaia parallaxes in principle allows the measurement of stellar masses without the use of stellar models \citep{stassun_empirical_2018}. \citet{bastien_granulation_2016} expanded their initial framework to use an iterative process to derive typical granulation timescales. \citet{pande_surface_2018} and \citet{bugnet_fliper_2018} expanded the approach by \citet{bastien_granulation_2016} to work in the frequency domain using Kepler stars. The large number of available space-based light curves also inspired data-driven and machine-learning approaches such as local-linear regression \citep{sayeed_swan_2021} and transformer-based models \citep{pan_astroconformer_2024}.

In addition to stars with convective envelopes, variability with observational signatures akin to granulation has now been observed in more massive stars. First evidence for this was presented by \citet{kallinger_evidence_2010} in $\delta$\,Scuti stars observed by MOST, and recent theoretical work suggests that stars without convective envelopes may indeed show signatures of granulation \citep{tschernitz_granulation_2025}. For more massive stars, TESS has revealed ubiquitous stochastic low-frequency variability \citep{bowman_low-frequency_2019}, which for some stars has been linked to subsurface convection \citep{lecoanet_low-frequency_2019} and/or internal gravity waves \citep{edelmann_three-dimensional_2019, horst_fully_2020, serebriakova_eso_2024}.  This new class of variability allows insights into mixing processes in massive stars \citep{varghese_numerical_2025}, which are the progenitors of neutron stars and black holes (see Section \ref{sec:bh}).

The observational breakthroughs in the study of convection from space-based photometry has significant implications for our understanding of stars. The remarkably simple scaling of granulation (and oscillation) with stellar properties stands in contrast to the complex nature of convection as a physical process, which requires three dimensional models to be described accurately. Modeling granulation and oscillation amplitudes from first principles is poorly understood owing to large uncertainties when modeling the convection that stochastically drives and damps these processes \citep{houdek_interaction_2015}, and attempts to connect models to observations often require ad-hoc correction factors \citep{cranmer_stellar_2014}. 
However, significant theoretical progress is being made \citep[e.g.][]{trampedach_improvements_2014, zhou_does_2024, eitner_m3dis_2024,witzke_testing_2024, perdomo_garcia_effect_2025} and space-based missions will continue to provide benchmarks to improve our understanding of convection.

\subsection{Rotation and Stellar Activity}

Rotation is among the most fundamental properties of stars. After inheriting angular momentum from the collapse of a protostellar cloud, stars  change their rotation periods due to astrophysical processes such as angular momentum loss due to stellar winds and tidal interactions with other stars and planets. The former leads to the powerful possibility of using rotation periods as a proxy for stellar age \citep[gyrochronology,][]{skumanich_time_1972, barnes_ages_2007}. Stellar activity is closely related to rotation, being driven by magnetic fields that are generated through internal dynamo processes \citep{charbonneau_dynamo_2010, brun_magnetism_2017}. Observational signatures include photospheric dark spots and bright faculae, flares and chromospheric emission. A review of stellar rotation and activity for main-sequence stars is given by \citet{santos_kepler_2024}. Probes of interior rotation through asteroseismology, which are not covered here, have been reviewed by \citet{aerts_asteroseismic_2024}.

\begin{marginnote}[]
\entry{Gyrochronology}{Method to determine stellar ages using the spin-down of stars with convective envelopes over time.}
\end{marginnote}

Space-based light curves are exquisitely suited to study stellar rotation and activity. The small spot sizes in typical Sun-like stars are inaccessible to ground-based observations. Space-based photometry from Kepler, K2 and TESS resulted in catalogs of rotation periods for tens of thousands of cool stars across the HR diagram \citep{mcquillan_rotation_2014, ceillier_surface_2017, santos_surface_2021, reinhold_stellar_2020, claytor_tess_2024}, including a large number of young stars \citep{rebull_rotation_2018, douglas_k2_2019}. These observations have yielded a range of surprises in our understanding of angular momentum transport, including weakened magnetic braking in Sun-like stars \citep{van_saders_weakened_2016, hall_weakened_2021}, stalled spin-down in K-dwarfs \citep[e.g.][]{curtis_temporary_2019}, and a ``gap'' in the rotation period distribution \citep{mcquillan_rotation_2014}. Space-based light curves have also yielded detections of apparent rotational modulations in unexpected parameter spaces such as A stars \citep{balona_starspots_2017}, which may be caused by chemical spots.

Space-based data also showed that stellar activity is remarkably well traced by the amplitude of rotational modulation. Kepler data demonstrated that more active stars tend to oscillate with lower amplitudes \citep{garcia_corot_2010, chaplin_evidence_2011}, which was later quantified through the suppression of oscillations in red giants in tight eclipsing binaries \citep{gaulme_red_2013, gaulme_surface_2014}. Activity measurements in thousands of Kepler main-sequence stars confirmed that stellar activity closely traces stellar rotation up to a certain threshold, after which the activity level saturates, and revealed interesting differences to spectroscopic indicators such as the absence of stars with intermediate activity levels \citep{santos_kepler_2024}.

A fascinating debate enabled by space-based photometry concerns the activity level of the Sun. Early Kepler results hinted that other Sun-like stars appear to be noisier than the Sun \citep{gilliland_kepler_2011, gilliland_kepler_2015}, which provided the argument for an extended Kepler mission (since more data was required to detect Earth-like transiting planets). Other studies concluded that the Sun does not appear to be more active than other Sun-like stars \citep{basri_photometric_2010, basri_comparison_2013}. The precise stellar parameters from Gaia renewed interest in this topic,  again concluding that stars that are nearly identical to the Sun are more variable \citep{reinhold_sun_2020, zhang_solar-type_2020}. This, however, may also be attributed to selecting stars with cooler temperatures and higher metallicities \citep{metcalfe_comment_2020, reinhold_reply_2020}. Updated rotation period and activity measurements conclude that the Sun appears to be consistent with other Sun-like stars \citep{mathur_magnetic_2023, santos_temporal_2023}. Possible explanations consistent with current literature include that the Sun is transitioning into a lower activity state \citep{,metcalfe_stellar_2016}, that the Sun and other Sun-like stars occasionally experience epochs of high activity \citep{reinhold_sun_2020}, or that the Kepler sample does not include a sufficient number of stars that are truly Sun-like \citep{herbst_comparing_2025}. 

\subsection{Flares and Outbursts} 

Stellar flares are sudden releases of energy on the surfaces of stars that occur on timescales of minutes, typically characterized by a steep flux increase followed by exponential decay. Understanding the frequency, duration, and energy distributions of flares for stars across the H-R diagram has wide-ranging implications in astrophysics, including the magnetic activity evolution of stars, the habitability of exoplanets, and high-energy space weather events on our own Sun that can affect the Earth.

\begin{marginnote}[]
\entry{Stellar Flares}{Sudden brightenings resulting from magnetic reconnection on stellar surfaces; probe magnetic activity and habitability environments.}
\end{marginnote}

The power of space-based time-domain data for the study of flares was realized as soon as the first quarter of 1-minute cadence Kepler data, which yielded clear detections in hundreds of cool M and K stars  \citep{walkowicz_white-light_2011}. Subsequent studies initially focused on M dwarfs, finding that more active stars flare more often, but also demonstrating a large range of flare frequency and energy that blurred the distinction between active and inactive stars through classical spectroscopic activity indicators \citep{hawley_kepler_2014}. Larger scale catalogs enabled detections of hundreds of thousands of flares in thousands of stars, demonstrating that flare activity decreases for low-mass field stars as their rotation rates decrease, and revealing a strong correlation with Rossby number, in line with expectations that stellar activity is coupled to flare rates \citep{davenport_kepler_2016}. Surprisingly, these studies also showed that while a single power law is unable to explain flare energies, the power law index is constant across spectral types, indicating a common physical mechanism \citep{davenport_kepler_2016, davenport_evolution_2019, van_doorsselaere_stellar_2017, yang_flare_2019}. Kepler also enabled the first detection of quasi-periodic pulsations in white light flares \citep{pugh_statistical_2016}.

\begin{figure}
\begin{center}
\includegraphics[width=5in]{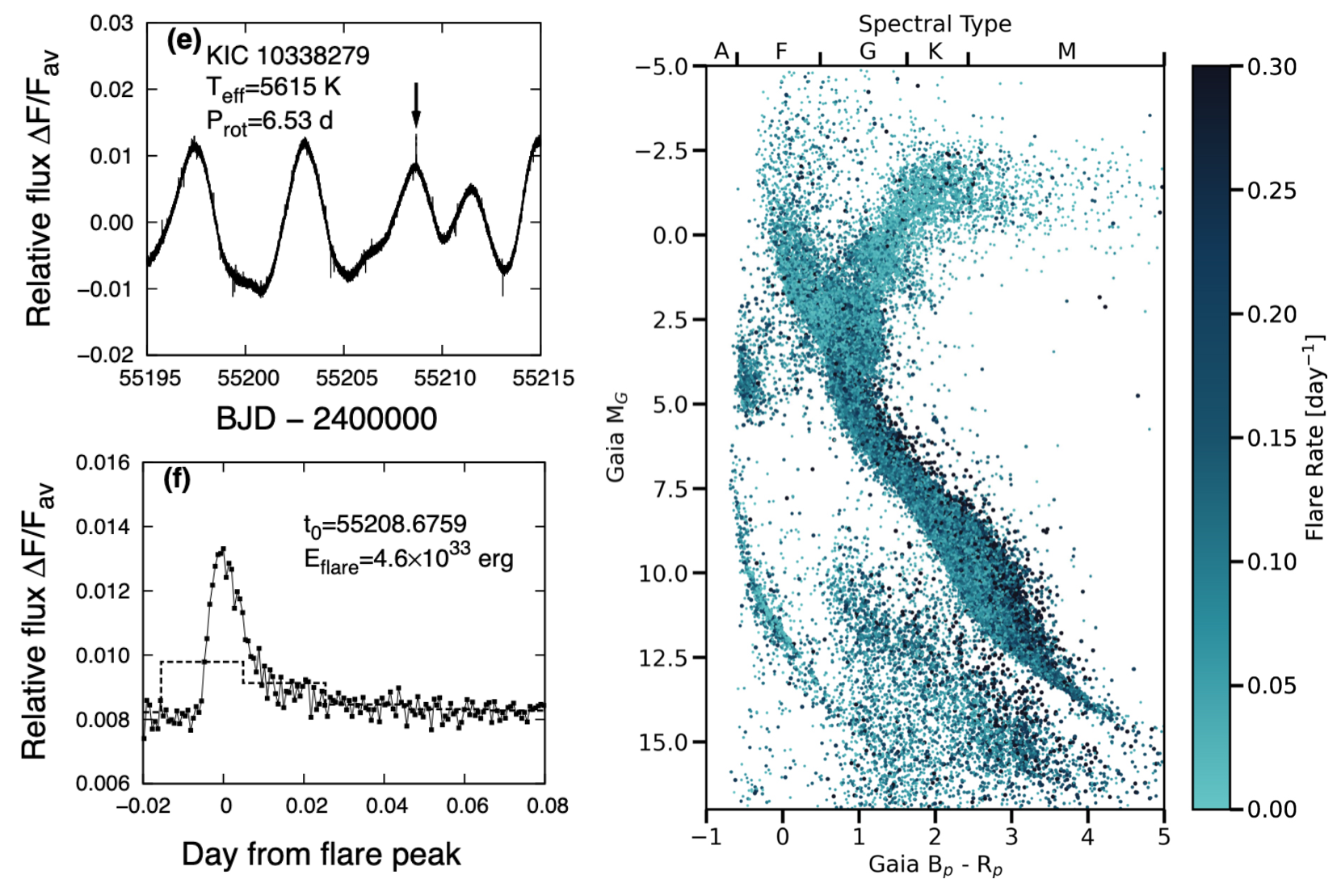}
\caption{\textbf{Space-based time-domain photometry has revolutionized our understanding of stellar flares.} Left, reproduced from \citet{maehara_statistical_2015}: Kepler light curve of a main-sequence G-type star showing a superflare. The bottom panel shows a zoom-in on the flare. Right, reproduced from \citet{feinstein_testing_2022}: Flare rates across the H-R diagram derived from TESS light curves.}
\label{fig:flares}
\end{center}
\end{figure}

Kepler light curves also enabled several surprising discoveries on flare properties of stars. An example is superflares on G-type stars \citep{maehara_superflares_2012}, which release flare energies an order of magnitude higher than those typically observed in the Sun (Figure \ref{fig:flares}, left). Follow-up studies demonstrated that superflares are correlated with large starspots, and surprisingly can occur for stars with rotation periods indicating a similar age to the Sun \citep{shibayama_superflares_2013, notsu_kepler_2019}. Another surprising discovery was that flares appear to be more wide-spread across the H-R diagram than previously thought, including A-type stars \citep{balona_flare_2015} and giant stars \citep{yang_flare_2019, ong_gasing_2024}. The observed flare properties of A stars appear to indicate a different mechanism than for stars with convective envelopes \citep{yang_flare_2019}.

The K2 Mission extended the success from Kepler by observing stellar flares across a wider range of spectral types. In particular, K2 provided first insights into the flare rates of ultracool dwarfs \citep{gizis_k2_2017}. K2 also opened up the study of flare rates in clusters with a wider range of ages than those available with Kepler \citep{ilin_flares_2019,ilin_flares_2021}, largely confirming that flare rates show a strong dependence on age, and that flare rates decline faster with age for higher mass stars. The community-driven target selection by K2  also enabled simultaneous multiwavelength campaigns, combining space-based optical observations with ground-based spectroscopy and space-based near-UV observations with Swift \citep{paudel_multiwavelength_2024}. These studies have laid important ground-work to understand the high-energy environment of low-mass stars, in particular ultracool dwarfs, which have become prime targets for detecting exoplanets in the habitable zones of stars.

The most recent milestone in flare studies has been enabled by the TESS Mission through two innovations: the unprecedented number of all-sky 2-minute cadence observations, and the introduction of 20-second cadence observations. The former enabled large-scale, all-sky catalogs of stellar flares, allowing an order of magnitude increase in stars with detected flares over Kepler and K2 \citep{gunther_stellar_2020, feinstein_flare_2020}. These studies solidified that the flare frequency distributions in stars appear to be universal across a large range of stellar masses, and that photometric variations with flare-like morphologies in observations appear to be ubiquitous across the H-R diagram (Figure \ref{fig:flares}, right). At least part of these detections are likely unrelated to classical flares and may instead be related to pulsational phenomena such as stochastic outbursts, which were first revealed through Kepler/K2 observations of white dwarfs \citep{bell_kic_2015, hermes_second_2015}. 

TESS 20-second cadence data provided spectacular new insights into the detailed morphology of flares, which was smeared out in most previous observations , and demonstrating that  quasi-periodic pulsations are common \citep{gilbert_flares_2022, howard_flaring_2022}. The detailed morphology of flares in M dwarfs revealed by TESS has led to connections with observations in the Sun, linking the observed multimodal variability to plasma condensation in the corona \citep{yang_possible_2023}. Fast cadence data from TESS has also been used to study the effect of star-planet interactions on stellar flares \citep{ilin_searching_2022}.

The long baselines of space-based time-domain have enabled the exploration of using flare variability as a tracer for stellar activity cycles \citep{scoggins_using_2019}. Evidence for such variability has been detected in TESS data spanning over five years \citep{feinstein_evolution_2024} and when combining observations of the same star from Kepler and TESS \citep{wainer_searching_2024}. As the TESS time baselines extend even further, flares could provide a powerful probe to measure long-term activity cycles of stars across the H-R diagram.

\subsection{Dippers, Bursters, and Exotic Variables}

A common theme in the space-based time-domain revolution is that astrophysical processes which previously appeared ``simple'' were revealed to have extraordinary complexity. One of the best examples is the variability in young stars (broadly defined as stars with ages $\lesssim$ 100\,Myr, with material from the forming disk typically disappearing within $\lesssim$ 10\,Myr) as well as the discovery of exotic variability in mature stars.

\begin{marginnote}[]
\entry{Young Stars}{Stars in early formation stages (ages $\lesssim$ 10-100\,Myr), showing variability from accretion, circumstellar obscuration, or disks.}
\end{marginnote}

\begin{figure}
\begin{center}
\includegraphics[width=5in]{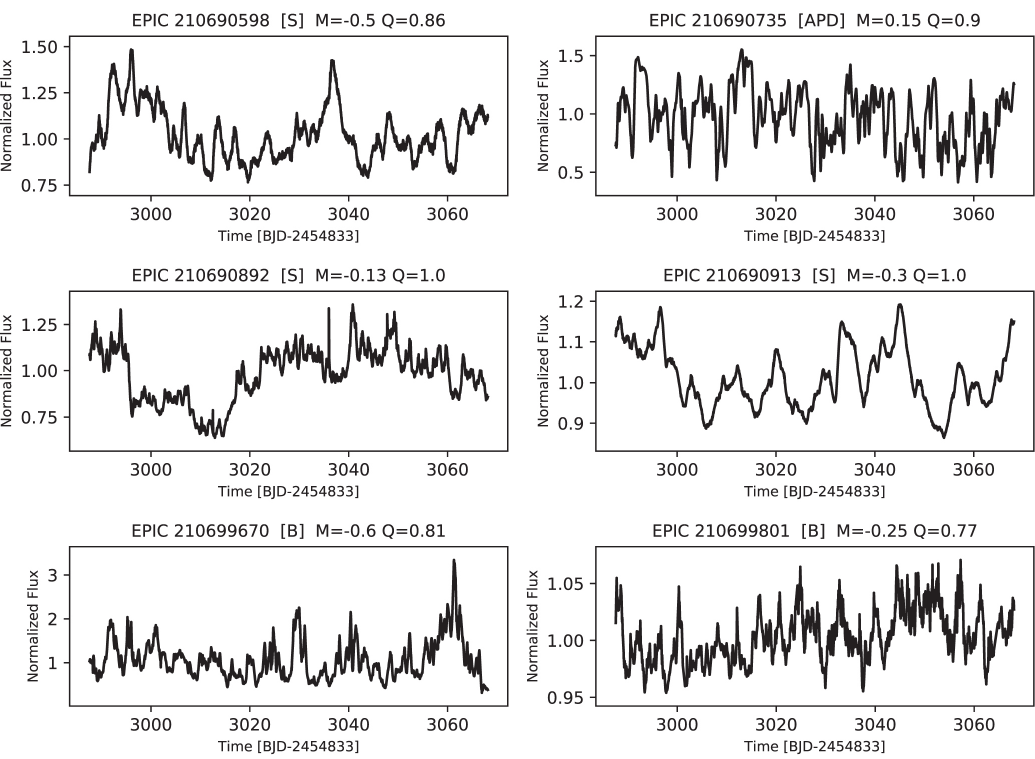}
\caption{\textbf{The remarkable variability of young stars revealed through space-based time-domain photometry}. From \citet{cody_many-faceted_2022}: each panels shows a 70-day light curve from K2 of disk-bearing young stars in the Taurus star-forming region. Abbreviations in square brackets denote the morphological classification scheme from \citet{cody_csi_2014}.}
\label{fig:yso}
\end{center}
\end{figure}

Time variability observations of young stellar objects have a long history \citep{joy_t_1945}. Early space-based results from CoRoT and Spitzer of the open cluster NGC\,2264 revealed a rich complexity of variability, broadly attributed to spots, obscuration events from circumstellar material that is left over from the formation, and accretion fluctuations \citep{cody_csi_2014}. The rich data quality motivated a classification scheme, separating symmetric variability, predominant decreases in brightness (the ``dippers''), and predominant increases in brightness (the ``bursters''). Combined with the periodic nature of the signal, these observations motivated a morphological classification scheme using periodicity, stochasticity, and symmetry to separate the light curves (Figure \ref{fig:yso}). 

Kepler provided few insights into the variability of young stars due to its target selection focusing on mature solar-type stars. A recent exception was the Gaia-based discovery of a young moving group that was serendipitously observed by Kepler \citep{bouma_kepler_2022}. What Kepler lacked was recovered by K2, which, thanks to its pointing strategy in the ecliptic plane, covered numerous young associations with high-precision space-based photometry for the first time. K2 revealed spectacular variability in young star-forming regions, including upper Scorpius and Ophiuchus \citep{ansdell_dipper_2016, cody_many-faceted_2018, hedges_discovery_2018}, the Lagoon Nebula Cluster \citep{venuti_multicolor_2021}, and Taurus \citep{cody_many-faceted_2022, roggero_dipper_2021}. These observations revealed that younger populations are more dominated by phenomena attributed to accretion (e.g. bursts), while older populations show a large fraction of dippers consistent with more evolved discs, implying that accretion timescales are shorter than disc dispersal timescales \citep{cody_many-faceted_2022}.

Light curves also probe geometric effects such as disc inclinations. Surprisingly, some dipper stars were found to have face-on orientations, and samples with isotropically orientated outer discs (resolved with sub-mm observations by ALMA) show that the dipper phenomenon is unrelated to the outer disc geometry \citep{ansdell_are_2020}. This implies that either dipper events may not be predominantly caused by dust in the inner disc, or that inner and outer disc misalignments may be common. \citet{cody_many-faceted_2022} found that inclination correlations exist only for periodic and quasiperiodic sources, while aperiodic sources appear largely isotropic, suggesting that different subclasses of dipper stars may be attributed to astrophysical phenomena. Overall, while young star variability is shaped by age and geometry, a consistent physical picture is still elusive and likely related to the complex interplay of accretion, magnetic fields, and disc evolution. New observations by TESS have already started to add valuable data to this topic \citep{tajiri_dippers_2020}, including the surprising discovery of dipper stars substantially older than the canonical $\approx$\,10\,Myr for which disks should have  dispersed \citep{gaidos_planetesimals_2022}.

Space-based time-domain data has also revealed new classes of exotic variable stars. An example is the ``complex periodic variables'', which show nearly periodic brightness decreases that are superimposed on rotational variability due to spots \citep{stauffer_orbiting_2017}. The phenomenon occurs mostly on young pre-main sequence M dwarfs without evidence of disks, and at times can undergo ``state-changes'' during a cycle. Combined data from K2 and TESS are now favoring a picture in which this variability is caused by gas or dust that is co-rotating with the star \citep{bouma_transient_2024}. 

One of the most remarkable aspects of the space-based time-domain photometry revolution is the discovery of exotic variability in mature, solar-type stars. The most famous example is Boyajian's star, a seemingly normal F-type main-sequence star demonstrating large, irregular brightness variations \citep{boyajian_planet_2016}. Other examples include the ``random transitter'' HD\,139139, a star exhibiting seemingly random transits without periodicities during the K2 observing window \citep{rappaport_random_2019} that may have disappeared at later epochs based on observations with CHEOPS \citep{alonso_no_2023}. Consensus explanations for the exotic variability remain elusive, and the continuation of space-photometry revolution is set to uncover more of these exotic results.

\subsection{Binary Stars} 

Binary stars are benchmark systems in astrophysics for determining fundamental properties of stars with no or little model dependence \citep[e.g.][]{torres_accurate_2010}. The traditional prototype are eclipsing binary stars, which for the special case of double-lined spectroscopic (SB2) binaries allow dynamical measurements of radii and masses of the binary components. Contact binaries are stars with close orbital periods for which the binary components are tidally deformed, yielding near sinusoidal light curve shapes. For binaries on wider orbits, astrometric and interferometric monitoring of binary orbits yields similar constraints. Binary stars with pulsating components provide unique laboratories for testing stellar physics (see Southworth \& Bowman, this volume).

Eclipsing and contact binaries are accessible through ground-based photometry, and time-domain surveys such as OGLE, ZTF, ASAS-SN, and ATLAS have yielded large catalogs of such binary star systems \citep{soszynski_ogle_2017, heinze_first_2018, chen_zwicky_2020, rowan_value-added_2023}. The high precision and continuous coverage of space-based time-domain missions revealed physical effects that were previously not detected. Prominent examples include the first detection of Doppler beaming, a periodic modulation of the photon emission rate as the stars move towards the observer \citep{bloemen_kepler_2011}, the discovery of ``heartbeat'' stars, binaries on eccentric orbits which cause dynamically induced tidal distortion and pulsations \citep{thompson_class_2012}, and the discovery of tidally-tilted pulsators \citep{handler_tidally_2020}. The last two studies are examples of the new field of ``tidal asteroseismology'' that was sparked by the precision of space-based time-domain photometry \citep{burkart_tidal_2012,guo_listening_2020}.

\begin{marginnote}[]
\entry{Heartbeat Stars}{Eccentric binaries showing tidally induced brightness variations and pulsations near periastron.}
\end{marginnote}

\begin{figure}
\begin{center}
\includegraphics[width=6in]{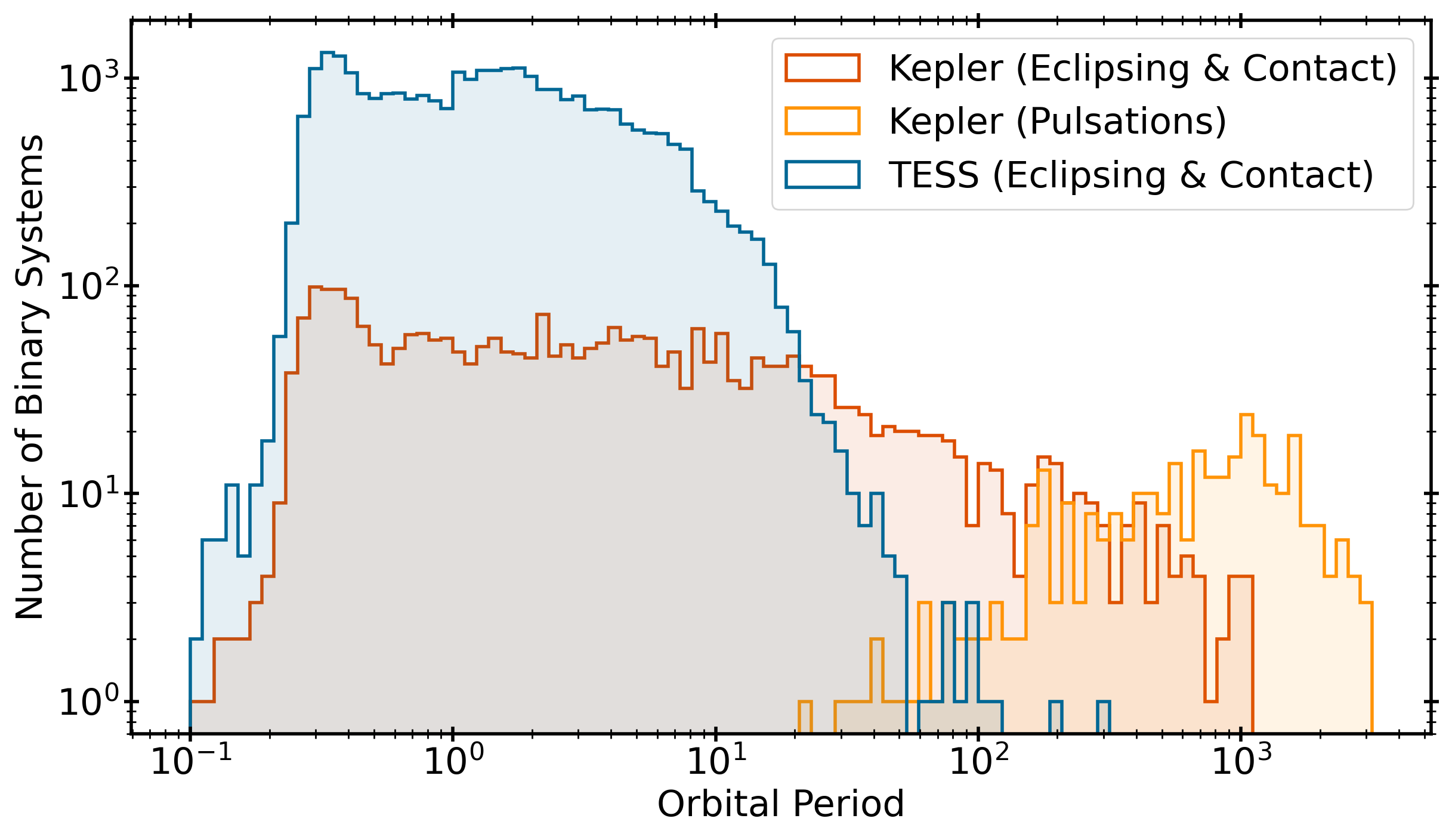}
\caption{\textbf{Space-based time-domain missions have uncovered tens of thousands of binaries with orbital periods spanning over five orders of magnitude.} Histogram of orbital periods of binaries discovered by Kepler and TESS, separated into contact/eclipsing binaries and binaries detected using pulsation timing. Data taken from \citet{kirk_kepler_2016}, \citet{murphy_finding_2018}, \citet{green_15_2023}, \citet{prsa_tess_2022}, and \citet{kostov_tess_2025}.}
\label{fig:binaries}
\end{center}
\end{figure}

Another breakthrough was the discovery of high-order eclipsing systems. The first examples from Kepler included hierarchical triply eclipsing triple systems with a main-sequence primary star \citep{carter_koi-126_2011} and an evolved red giant primary star \citep{derekas_hd_2011}, the latter also showing tidally forced variability from the orbit of the lower mass binary. Eclipse timing variations using Kepler data revealed evidence for hundreds of triple systems, providing statistical constraints on the distributions of orbital parameters in higher order binaries \citep{borkovits_comprehensive_2016}. The large number of stars monitored by TESS has enabled the discovery of even higher order systems, including over a hundred eclipsing quadruple systems \citep{kostov_101_2024}, the first known sextuple system with three eclipsing binaries \citep{powell_tic_2021}, and a triply eclipsing sextuple star system around a previously known eclipsing binary star \citep{zasche_detection_2023}. Space-based missions have even yielded discoveries of new eclipsing binaries around naked-eye stars that have previously been missed by ground-based observations \citep{bedding_dance_2019, hey_parameters_2022}.

A legacy product of the space-based time-domain missions is large catalogs of binaries with measured orbital parameters. The nominal Kepler mission discovered over 3000 eclipsing binaries \citep{prsa_kepler_2011, kirk_kepler_2016}, followed by several hundred systems so far discovered in the K2 dataset \citep{armstrong_k2_2015, maxted_discovery_2018}. A novel method to detect binaries discovered with Kepler was the timing of coherent pulsations, allowing measurements equivalent to radial velocities through frequency \citep{kurtz_validation_2015} and phase \citep{murphy_finding_2014} modulations. The signal strength of pulsation timing increases with orbital period \citep{hey_forward_2020}, thus providing a complementary method to eclipsing binaries. Finally, the TESS mission has so far delivered over 10,000 contact binary candidates on close orbits \citep{green_15_2023} and thousands of detached eclipsing binaries \citep{prsa_tess_2022,ijspeert_statistical_2024, kostov_tess_2025}. Overall, the space-based time-domain missions have so far discovered tens of thousands of new binaries with orbital periods spanning nearly five orders of magnitude (Figure \ref{fig:binaries}).

Space-based telescopes also advanced our understanding of cataclysmic variables (CVs). CVs consist of a white dwarf primary which is accreting material from a donor star, leading to both irregular and periodic brightness variations. Kepler data allowed the discovery of new CVs \citep{howell_spectroscopy_2013}, including examples showing eclipses \citep{scaringi_kepler_2013}. The continuous coverage also enabled the confirmation that periods that are slightly different than the orbital periods (so-called ``superhumps'') can be attributed to hot spots in a tilted, precessing disk surrounding the white dwarf \citep{gies_kic_2013} and led to novel insights in the related phenomenon of superoutbursts \citep{osaki_cause_2013}. K2 and TESS increased the sample of CVs with space-based light curves to a few dozen, revealing hitherto unknown variability \citep{bruch_tess_2022}. CVs are an excellent example of stellar astrophysics where the precision and continuity of space-based light curves led to new insights even for variable stars with large amplitudes that have been studied from the ground for decades.

\section{Galactic Astronomy} 

\subsection{Galactic Archaeology}
\label{sec:ga}

The study of the formation and evolution of the Milky Way has emerged as a key topic in astrophysics over the past decades \citep[see][for a review]{helmi_streams_2020}. Historically, the study of stellar populations has been driven by large-scale photometric and astrometric datasets such as the Geneva-Copenhagen survey \citep{nordstrom_geneva-copenhagen_2004} and Hipparcos \citep{van_leeuwen_validation_2007}. Over the past decades, multiplexed spectroscopic surveys have added high-precision chemical abundances, which are associated with the birth environment \citep[``chemical tagging'', see][]{freeman_new_2002}, while the Gaia mission provided exquisite astrometry for billions of stars \citep{gaia_collaboration_gaia_2023}. 

The major contribution of space-based time-domain missions was to measure ages of stars on sufficiently large scales to provide a chronological order of galactic stellar populations and events - a subfield now commonly referred to as ``galactic archaeology''. Since more luminous stars probe larger distances, the primary tool to provide this information is asteroseismology of red giants.  The mass, metallicity, and internal physics of a red giant star directly define its age, and thus the combination of space-based light curves and spectroscopic ground-based data thus provides the possibility to measure ages for a significant number of stars on galactic scales. 

\begin{marginnote}[]
\entry{Galactic Archaeology}{Reconstruction of the Milky Way’s history using stellar ages, chemistry, and kinematics.}
\end{marginnote}

The potential of asteroseismology for galactic archaeology was first realized by \citet{miglio_probing_2009}, who observed qualitative agreement between population synthesis models and asteroseismic detections in 800 stars observed by CoRoT. Later work analyzing multiple CoRoT pointings revealed significant differences in mass distributions, most likely due to different vertical scale heights being probed \citep{miglio_differential_2013}. This was also observed in a sample of Kepler dwarfs and subgiant stars \citep{chaplin_ensemble_2011}. The large Kepler red giant sample provided  strong potential and challenges for advancing galactic archaeology, the latter being predominantly due to the poorly understood selection function of red giants \citep{sharma_stellar_2016}. Building upon this lesson, the K2 galactic archaeology program \citep{stello_k2_2017} was the first large-scale coordinated effort to use space-based light curves as probes of galactic structure \citep{zinn_k2_2020, zinn_k2_2022}. Finally, the full-frame images from TESS eliminated the need for target selection, yielding a near all-sky view for nearby oscillating red giants free of selection biases \citep{mackereth_prospects_2021, hon_quick_2021}. 

\begin{figure}
\begin{center}
\includegraphics[width=4.5in]{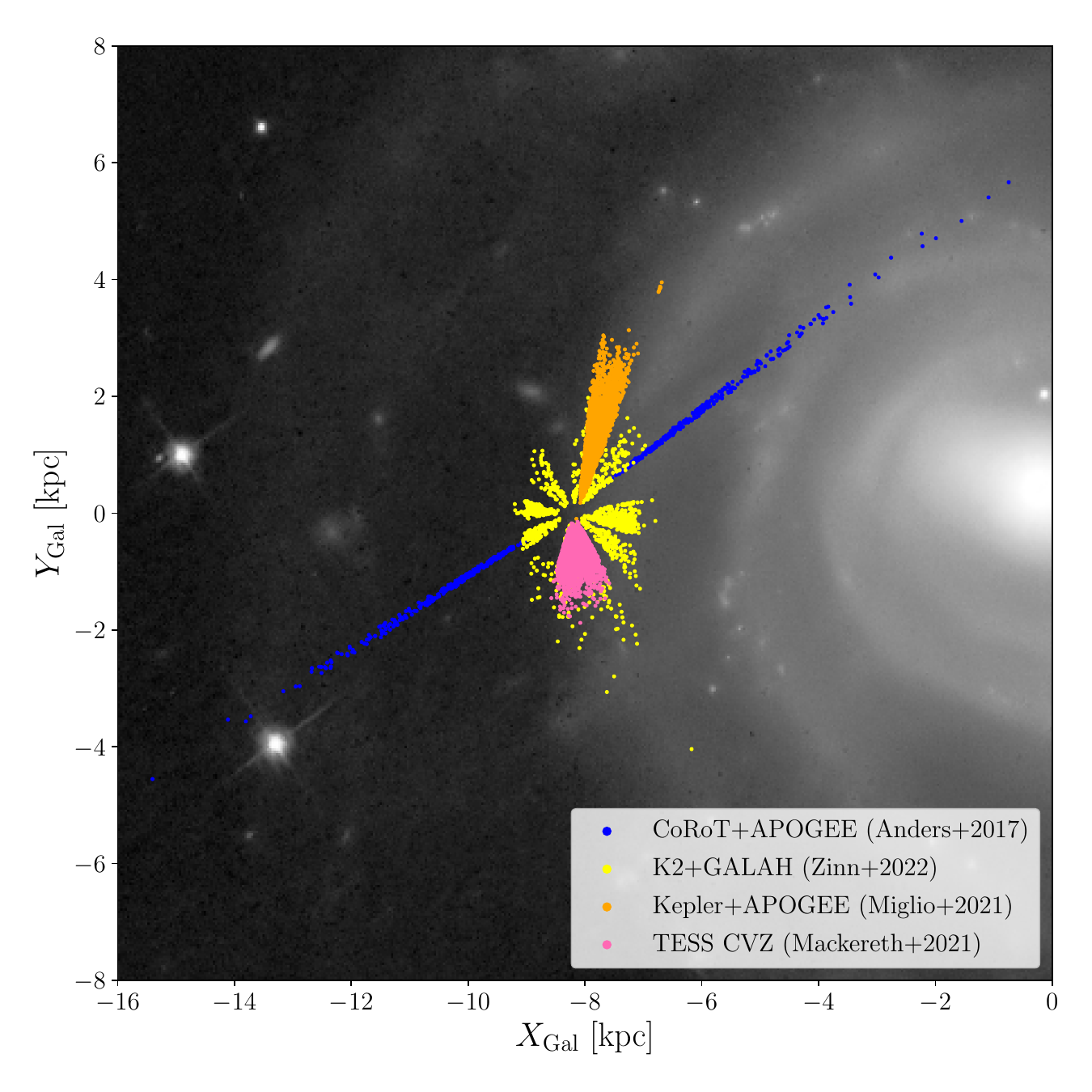}
\caption{\textbf{Space-based photometry has provided a new tool to study old stellar populations in our galaxy.} Top down view of an HST image of the Milky Way analog UGC 12158. Colored symbols show stars for which space-based light curves from different missions have yielded asteroseismic detections and for which spectroscopic follow-up data exists. Image credit: Friedrich Anders, based on work from \citet{anders_spectroscopic_2023}.}
\label{fig:ga}
\end{center}
\end{figure}

In parallel, spectroscopic surveys specifically targeted stars with space-based detections of oscillations to provide complementary chemical abundances, which are critical to infer ages from asteroseismic detections. This includes APOGEE observations of the CoRoT \citep{anders_galactic_2017}, Kepler \citep{pinsonneault_apokasc_2014, serenelli_first_2017}, and K2 \citep{schonhut-stasik_apo-k2_2024} fields, AAT/HERMES observations of K2 and TESS stars \citep{sharma_tess-hermes_2018, sharma_k2-hermes_2019}, LAMOST observations of predominantly the Kepler/K2 fields \citep{zong_phase_2020}, and observations of the recently commissioned 4MOST telescope \citep{bensby_4most_2019}. Combined, these samples probe substantial areas of our galaxy (Figure \ref{fig:ga}).

Ages provided by space-based photometry yielded expected results and surprising discoveries. \citet{silva_aguirre_confirming_2018} used Kepler data to show that stars enriched in $\alpha$-process elements relative to iron are on average $\approx$\,8\,Gyr older than stars that are $\alpha$ poor, consistent with galactic chemical evolution models and distinct ``thin'' and ``thick'' disk populations in the Milky Way. \citet{chiappini_young_2015} and \citet{martig_young_2015} identified an unexpected population of young alpha-rich stars, which may be the result binary interactions which strip off mass from the primary star, thus yielding a lower mass and (seemingly) older age \citep{grisoni_k2_2024}. Combining ages with radial metallicity gradients and kinematics support a picture in which stars in the galaxy undergo radial migration \citep{anders_galactic_2017, sharma_fundamental_2021} and that the galaxy evolved ``inside-out'' and may have experienced a late burst of star formation \citep{warfield_apo-k2_2024}, confirming earlier results from photometric and spectroscopic surveys \citep{casagrande_new_2011, bensby_exploring_2014, bergemann_gaia-eso_2014}.


Space-based photometry also provided first age estimates for non-disk components of our galaxy. Specifically, determining the age of a star that formed outside our galaxy puts a lower limit on the time that the star joined our Milky Way (e.g. a merger event). Kepler and TESS data showed that Gaia-Enceladus, a major merger of the Milky Way that is now commonly associated with the ``birth'' of the kinematic thick disk, occurred at least $\approx$\,8 years ago \citep{chaplin_age_2020, montalban_chronologically_2021, grunblatt_age-dating_2021}. Furthermore, the dwarf galaxy merger that led to the Helmi stream occurred at least $\approx$\,12 Gyr ago \citep{lindsay_precise_2025}. 

TESS observations also detected oscillations in ``halo'' stars \citep{borre_age_2022}, the oldest component of our galaxy whose origin is still uncertain. A challenge is that reliable ages for metal-poor stars require modeling of individual frequencies, which has only been done for a small number of very-metal poor stars \citep{huber_stellar_2024, larsen_pushing_2025} and becomes challenging for more luminous giants. 
The oldest ages of stars in the Milky Way may be used to constrain cosmological models and tensions between local and distant measures of the Hubble constant \citep{moresco_measuring_2024}. Improvements to stellar model physics will be required to assess the accuracy of such estimates. Achieving this goal is a major motivation for the planned HAYDN Mission \citep[][see Section \ref{ch:future}]{miglio_haydn_2021}.

A side product of the space-based revolution of galactic archaeology has been constraints on interstellar dust. This is enabled by the combination of high-precision spectroscopic parameters and asteroseismic surface gravities, which pin down the spectral energy distribution and allow extinction to be fitted as a free parameter when modeling broadband photometry. Application of this method to  Kepler data \citep{rodrigues_bayesian_2014} yielded strong systematic differences to traditional 3D dust maps. Future extensions of spectroscopic surveys with surface gravities calibrated by asteroseismology \citep[e.g.][]{godoy-rivera_testing_2021} will provide the potential to constrain large-scale empirical dust maps.

\subsection{Neutron Stars and Black Holes} 
\label{sec:bh}

The Milky Way is predicted to be home to millions of solar-mass black holes, several thousand of which are expected to reside in non-interacting binary systems \citep[e.g.][]{shao_population_2019}. Determining the demographics of such compact ``dark'' objects in binaries is critical for our understanding of massive star evolution and the progenitors of gravitational wave events. While several hundred black holes have been detected as X-ray sources in interacting binaries, the search for detached systems has significantly increased thanks to large scale surveys. Currently only a handful of such black holes are known in the field, all of which have been identified and confirmed through the combination of Gaia astrometry and radial velocities \citep{chakrabarti_noninteracting_2023, el-badry_red_2023, el-badry_sun-like_2023}. 

Neutron stars and black holes can be detected photometrically either through ellipsoidal variations (for short orbital periods), gravitational lensing as the as the compact object moves in front of the luminous companion (self-lensing binaries), or pulsation timing. The first method has yielded several candidate black hole detections from ground-based photometric surveys such as ASAS-SN (\citep{thompson_noninteracting_2019, jayasinghe_unicorn_2021, jayasinghe_giraffe_2022}. However, many of these could not be independently confirmed and are often compatible with non-single star evolution and binary mass transfer scenarios \citep[e.g.][]{el-badry_unicorns_2022}.

Space-based photometry holds strong promise to increase the known population of detached neutron stars and black holes. Kepler enabled the first discovery of a self-lensing binary star including a white dwarf and a main-sequence star \citep{kruse_koi-3278_2014}, with additional systems discovered later on \citep{kawahara_discovery_2018, masuda_self-lensing_2019}. First systematic searches using all-sky data from TESS have started to constrain the occurrence rate of white dwarfs around Sun-like stars to $\approx$1\% by looking for self-lensing events \citep{yamaguchi_search_2024}, and constrained the occurrence of short-period black holes around Sun-like stars to less than 10$^{-5}$ through a systematic search for ellipsoidal variability \citep{green_upper_2025}. Future searches, using extended TESS light curves, may yield the first photometric detection of nearby neutron stars and black holes in detached binaries.

Space-based data can also provide important complementary data to characterize neutron stars and black holes, for example by vetting candidates identified from spectroscopy \citep{jayasinghe_search_2023} and providing additional data to model ellipsoidal variations for candidates \citep{jayasinghe_giraffe_2022, jayasinghe_unicorn_2021}. TESS light curves have also enabled asteroseismology of red-giant stars with black holes detected using Gaia astrometry, providing precise characterization of the primary stars which ``host'' the dark companions \citep{hey_asteroseismology_2025}.
\section{Extragalactic Science} 

\subsection{Explosive Transients} 

Transient phenomena produce variability over limited time spans and, in general, do not show repeatable behavior. The study of explosive transients such as supernovae has wide-ranging implications in astrophysics, from constraining the expansion history of our universe by using Type Ia supernovae as standard candles to understanding the chemical enrichment of galaxies through core-collapse (Type II) supernovae. Recent reviews of explosive transients can be found in \citet{burrows_core-collapse_2021} and \citet{ruiter_type_2025}.

\begin{marginnote}[]
\entry{Explosive Transients}{Rapidly evolving events such as supernovae or kilonovae, probing progenitors and explosion physics.}
\end{marginnote}

The detection of supernovae has motivated ground-based imaging surveys such as the Palomar Transient Factory \citep[PTF,][]{law_palomar_2009}, the Zwicky Transient Facility \citep[ZTF,][]{bellm_zwicky_2019}, the All-Sky Automated Survey for Supernovae \citep[ASAS-SN,][]{shappee_man_2014}, and the recently commissioned Rubin Observatory Legacy Survey of Space and Time \citep{ivezic_lsst_2019}. Time-domain surveys to detect near-Earth asteroids and other rapid phenomena such as PanSTARRS \citep{chambers_pan-starrs1_2016} and ATLAS \citep{tonry_atlas_2018} also detect supernovae on a regular basis. All these surveys provide photometric coverage of large areas (and in the case of ASAS-SN and ATLAS the entire sky) every few nights on average.  Space-based time-domain missions such as Kepler/K2 and TESS provide complementary data to these surveys with faster cadence and better photometric precision, at the cost of a more limited survey area.

The first detection of Type Ia supernovae with Kepler was enabled through a Guest Observer program targeting hundreds of galaxies in the Kepler field \citep{olling_no_2015}. The resulting light curves showed exquisite precision compared to available ground-based data. The light curves revealed no evidence for the interaction of supernova ejecta with a stellar companion, solidifying the now established theory that type Ia supernova progenitors rarely contain non-compact objects and predominantly result from white dwarf binaries. In contrast, Kepler results on two core-collapse supernovae resulted in evidence for shock waves that interact with pre-existing material from a red supergiant progenitor \citep{garnavich_shock_2016}. Both findings were made possible by the early detection of the events thanks to continuous time-domain monitoring from space.

\begin{figure}
\begin{center}
\includegraphics[width=5in]{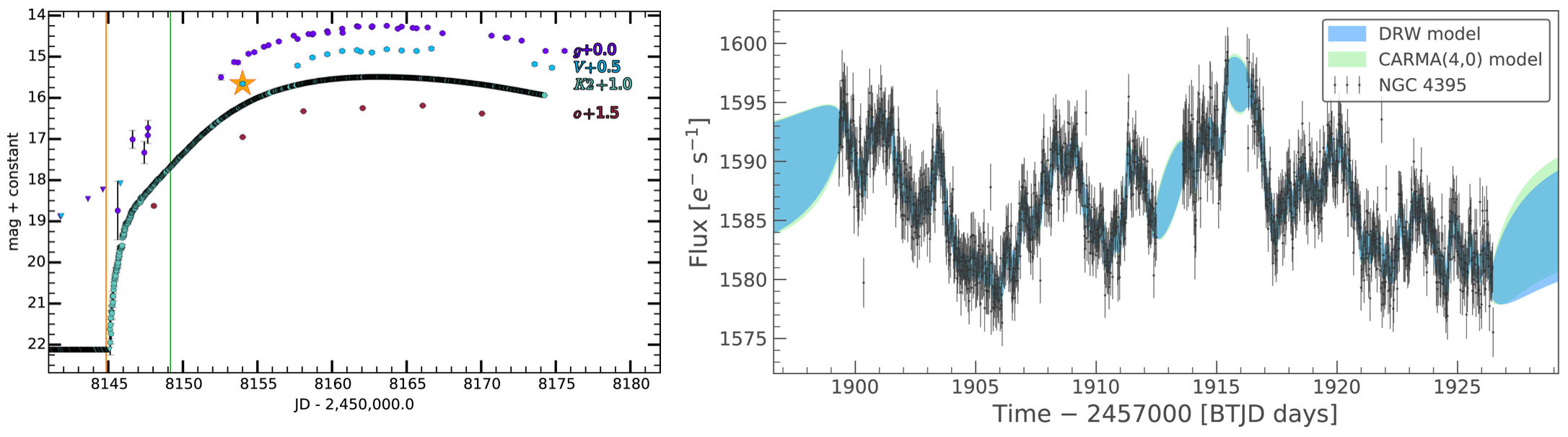}
\caption{\textbf{Space-based time-domain missions have provided exquisite new insights into the optical behaviour of extragalactic sources.} Left, reproduced from \citet{shappee_seeing_2019}: K2 light curve of a Type Ia supernova with a double-peaked rise component. Right, reproduced from \citet{burke_optical_2020}: TESS light curve showing the stochastic variability caused by accretion processes onto a supermassive black hole in the center of a dwarf galaxy. Colors show a damped random walk (DRW) and Gaussian Process (CARMA) model.}
\label{fig:sn}
\end{center}
\end{figure}

The success by Kepler led to systematic programs to search for explosive transients using the K2 Mission, which provided more extended sky coverage along the ecliptic. \citet{shappee_seeing_2019} and \citet{dimitriadis_k2_2019} used the early-time data from K2 to discover two components in the rise morphology of a type Ia supernova  (Figure \ref{fig:sn}), indicating that two distinct physical processes are at play during the explosion. Other Type Ia supernovae observed with similar precision were found to be consistent with a single power law rise \citep{wang_sn_2021}, highlighting the complexity and diversity in ``normal'' supernovae. K2 data also yielded the discovery of fast-evolving, luminous transients that may be the result of an explosion hitting material left behind from a pre-explosion mass-loss episode \citep{rest_fast-evolving_2018}. Finally, K2 allowed constraints on the progenitor of Type IIb supernova through high-cadence observations of the shock cooling light curve \citep{armstrong_sn2017jgh_2021}.

The large field of view of the TESS Mission (covering $\approx$5\% of the entire sky at a given pointing) further strengthened the synergy between supernova science and space-based time-domain missions. TESS data was used to characterize unusual Type Ia supernovae with emission features that may originate from circumstellar material \citep{vallely_asassn-18tb_2019} and confirm the flux excess in early-rise light curves of Type Ia explosions, with data defying current models \citep{wang_flight_2024}. TESS has allowed the first population study of Type Ia supernova characteristics using early-time light curves, hinting at two distinct types of Type Ia supernova populations based on the broad distribution of power law slopes \citep{fausnaugh_early-time_2021, fausnaugh_four_2023}. TESS also obtained early light curves of several core-collapse supernovae \citep{vallely_asassn-18tb_2019},  enabling more constraints on progenitors through shock-cooling \citep{wang_revealing_2023} and the detections of optical flashes and afterglows of gamma-ray bursts associated with core-collapse supernovae \citep{jayaraman_gamma-ray_2024, roxburgh_comprehensive_2024}. Finally, TESS is predicted to be able to provide optical counterpart measurements of kilonovae, transients caused by mergers of neutron stars and/or black holes that are sources of gravitational wave events \citep{mo_searching_2023}. TESS is now the most prolific optical space-based time-domain telescope for studying explosive transients, leading to it sometimes being referred to as the ``Transient Explorer Survey Satellite''\footnote{Originally coined by Ben Shappee and Jen van Saders (University of Hawai`i) as part of a TESS Guest Investigator Proposal.}.

\subsection{Active Galactic Nuclei} 

The centers of most galaxies are home to supermassive black holes. Accretion of material onto these black holes produces variability across the electromagnetic spectrum, ranging from radio emission from electrons accelerated along magnetic field lines, X-rays due to hot gas close to the black hole and optical/IR radiation in the accretion disk. Galaxies with central black holes masses $\gtrsim 10^5 M_{\odot}$ show luminosities that dominate the galaxy spectrum and are generally referred to as ``active galactic nuclei'' \citep[AGN, see][for a review]{netzer_revisiting_2015}.

\begin{marginnote}[]
\entry{Active Galactic Nuclei (AGN)}{Galaxies with accreting supermassive black holes producing emission across a broad range of the electromagnetic spectrum.}
\end{marginnote}

Time-domain variability of AGN can link the emission sources to physical processes in the accretion disk. 
The first space-based optical studies of AGN using Kepler data showed a wide variety of variability previously unseen in ground-based data \citep{mushotzky_kepler_2011, edelson_discovery_2014, revalski_investigating_2014, smith_evidence_2018}. The power spectrum slopes of an AGN sample showed correlations with bolometric luminosity and yielded the first detection of optical quasi-periodic oscillations, which are commonly detected in X-ray observations of stellar-mass black hole binaries, supporting the idea that accretion behaves similarly across a large range of masses \citep{smith_evidence_2018}. The characteristic frequency for the optical variability was $\approx$40 days and provided hints at a relationship between the variability timescale and black hole mass. Differences to the power slopes found in X-rays indicated that the optical variability is not just due to reprocessing of X-ray photons. Kepler also detected optical variability of a blazar, a subgroup of AGN with powerful radio emission due to jets that oriented towards the observer \citep{edelson_kepler_2013}.

K2 observations expanded the study of optical variability to dozens of blazars.
The results showed variability that is consistent with turbulence in the relativistic jet and accretion disk \citep{wehrle_measuring_2019, carini_measuring_2020}, and hinted at differences between two subclasses of blazars based on their radio spectrum variability \citep{wehrle_k2_2023}. K2 also provided the first opportunity for repeated observations  of blazars using space-based photometry, showing significant differences in power spectrum slopes over time \citep{wehrle_k2_2023}.

TESS enabled the combination of optical space-based photometry with simultaneous observations at other wavelengths. \citet{weaver_multiwavelength_2020} studied multi-wavelength variability of prototype blazar BL Lac, finding remarkable similarity in variability across optical, UV, and X-ray wavelengths. The data revealed frequency-dependent time lags relative to TESS, which can be explained by a model involving energization of the radiating electrons at a shock front and provided first constraints on the strength of the involved magnetic field. Unlike earlier K2 results, TESS data showed no difference in optical variability properties of blazar subclasses, despite expectations that one class (Flat Spectrum Radio Quasars, FSRQs) has a substantial optical contribution from the accretion disk. Characteristic optical timescales were found to range from $\approx$\,0.8-8 days, consistent with optical variability originating in the jet \citep{dingler_optical_2024}. 

TESS has also enabled the study of AGN in lower-mass galaxies. The characteristic variability timescale of a dwarf AGN was found to be significantly shorter than the corresponding timescales for higher-mass objects, providing strong evidence that the variability timescales of accretion disks correlate with black hole mass \citep[Figure \ref{fig:sn},][]{burke_optical_2020}. A consequence of this result is that TESS can also be used to identify AGN from light curve variability alone: \citet{treiber_revealing_2023} identified dozens of new AGN candidates based on TESS data and demonstrated that traditional selection criteria based on spectroscopy would fail for $\approx$\,40\% of the identified cases.

\subsection{Tidal Disruption Events}

When stars orbit too close to a supermassive black hole, tidal forces can exceed self-gravity, leading to a tidal disruption of the star. Remnants of the star can form an accretion disk around the black hole, leading to a luminous transient event. Such tidal disruption events (TDEs) provide insights into both black hole physics and stellar evolution, as well as identify dormant black holes that are otherwise undetectable \citep[for a recent review see][]{gezari_tidal_2021}.

\begin{marginnote}[]
\entry{Tidal Disruption Events}{Stellar disruption by a supermassive black hole, producing (at times periodic) flares and variability.}
\end{marginnote}

Space-based time-domain photometry has provided exquisite new insights into the early time variability of TDEs that are discovered with ground-based surveys. \citet{holoien_discovery_2019} presented the first TESS light curve of a TDE, allowing constraints on the time of first light and peak brightness, as well as providing constraints on the expansion velocities. TESS provided important constraints on the first periodic TDE, which is believed to be the result of a partial disruption event \citep{payne_asassn-14ko_2021}. The exquisite precision of TESS revealed subtle changes from one flare event to the other, which may be linked due to systematic differences in outbursts in each orbit \citep{payne_chandra_2023}. Finally, TESS has also been used to shine light on the nature of ambiguous nuclear transients by providing early-time (pre-transient) light curves that can distinguish between TDEs and AGN by searching for stochastic variability that is more consistent with AGN \citep{hinkle_tess_2023}.

\section{Solar System Science} 

\subsection{Small Bodies} 

The study of small solar-system bodies such as asteroids and comets informs our understanding of the formation of our solar system, such as the delivery of water to inner terrestrial planets and the motion of giant planets \cite[e.g.][]{morbidelli_source_2000}. Space-based time-domain photometry contributes to our understanding of small body populations by detecting light variations from rotationally modulated reflected sunlight. Rotation period and amplitude distributions give insight into the compositions of small bodies because for a given composition, asteroids reach a critical maximum rotation period beyond which they disintegrate. The composition in turn yields clues on their origin in the solar system.

\begin{marginnote}[]
\entry{Small bodies}{Objects orbiting the Sun that are not planets, dwarf planets, or natural satellites (moons). These include comets, asteroids, and interstellar objects.}
\end{marginnote}

The necessity for the K2 Mission to point along the ecliptic plane was fortuitous for solar system science. Early studies used asteroid data serendipitously collected within large superstamps that were downloaded to study clusters, gravitational microlensing events, and solar system planets \citep{szabo_uninterrupted_2016, podlewska-gaca_determination_2021}. K2 measured rotation rates of hundreds of main-belt asteroids, finding significantly longer rotation periods than ground-based surveys. This was interpreted as evidence that period distributions from such surveys are biased to short periods \citep{molnar_main-belt_2018}, since ground-based observations are hampered by day/night cycles and weather. 

K2 also obtained dedicated observations of Jovian Trojan asteroids. Early results constrained the binary fraction of Jovian Trojans to $\approx$\,20\%, and a rotation period distribution with a spin barrier that implies a composition consistent with comets, favoring an outer solar system origin \citep{szabo_heart_2017, ryan_trojan_2017}. Later studies increased the sample size to over 100 Trojans \citep{kalup_101_2021}, revealing a bimodality in their rotation distribution. Similar results of bimodal rotation periods were also found for Centaurs, a population of small bodies with orbits between Jupiter and Neptune \citep{marton_light_2020}.

Reaching into the outer solar system, K2's sensitivity allowed studies of some distant trans-Neptunian objects (TNOs). The analysis of 2007 OR1, the second most distant known TNO, yielded a significantly larger size and slower rotation rate than previously thought, implying a larger surface gravity and thus more likely retention of volatiles \citep{pal_large_2016}. \citet{kecskemethy_light_2023} analyzed a sample of 30 TNOs with K2 and found slower rotation and higher amplitudes for large TNOs compared to main belt asteroids. A possible explanation is that the compositions of TNOs allow the retention of irregular shapes at larger sizes than their inner solar system counterparts.

\begin{figure}
\begin{center}
\includegraphics[width=5in]{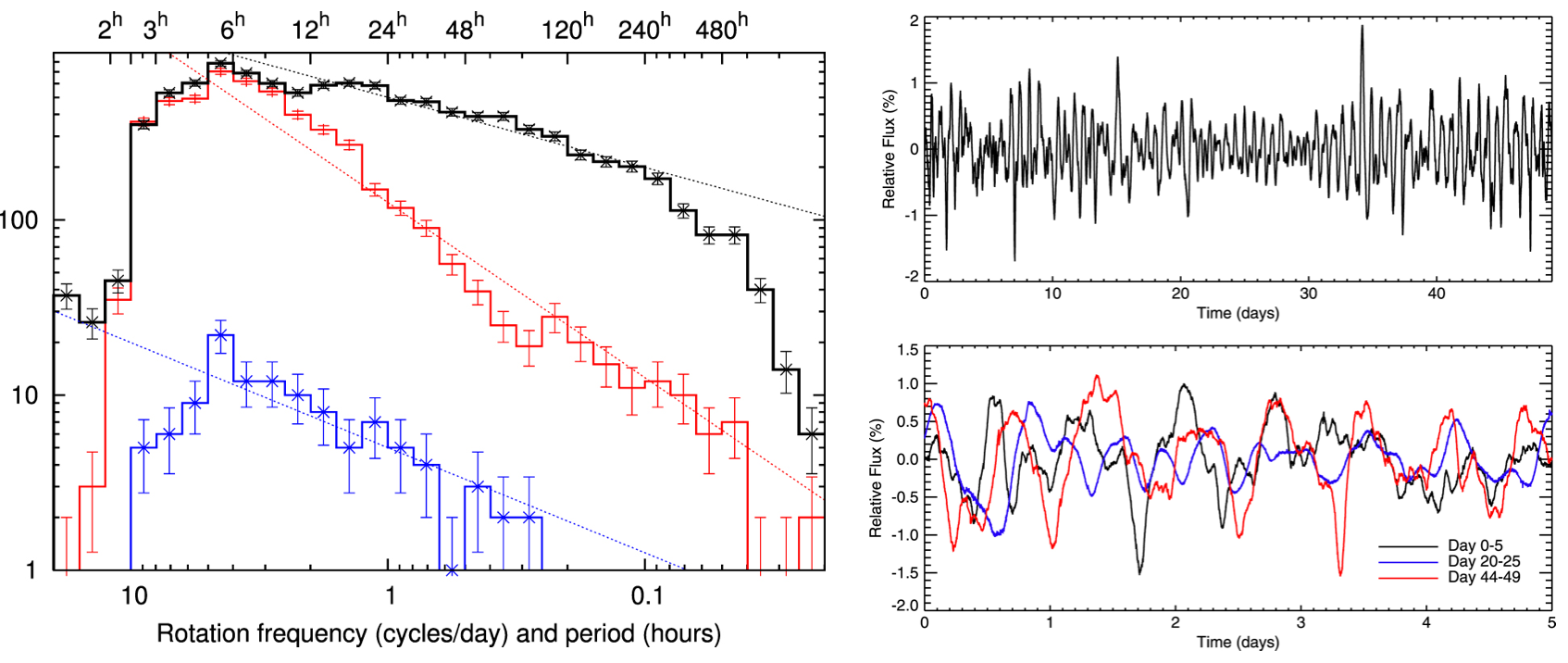}
\caption{\textbf{K2 and TESS data have provided new insights into the origin of small bodies and giant planets in the solar system.} Left, reproduced from \citet{pal_solar_2020}: Distribution of rotation periods of main-belt asteroids from previous surveys (red) and TESS (black), demonstrating that many long-period asteroids have previously been missed. Right, reproduced from \citet{simon_neptunes_2016}: K2 observations of Neptune, showing both the full continuous light curve (top) and 5 day segments with rapidly evolving variability due to clouds (bottom).}
\label{fig:solar}
\end{center}
\end{figure}

TESS has continued the success story by measuring rotation periods and amplitudes for thousands of main-belt asteroids, showing that long rotation periods were underestimated by previous ground-based surveys by up to an order of magnitude \citep[][Figure \ref{fig:solar}]{pal_solar_2020}. TESS also made discoveries that some small bodies rotate faster than expected: this includes fast-spinning medium-sized Hilda asteroids \citep{takacs_three_2025} and fast-rotating Jovian trojans \citep{kiss_three_2025}. These results imply either higher density compositions than previously assumed, collisions, or that the small body populations are more highly integrated with each other than previously thought.

A new class of solar system small bodies are interstellar objects, which are characterized by hyperbolic orbits. Three interstellar objects are currently known, discovered either by the ground-based surveys PanSTARRS \citep{chambers_pan-starrs1_2016, meech_brief_2017} and ATLAS \citep{tonry_atlas_2018, seligman_discovery_2025} or amateur astronomers. TESS data was used to obtain pre-discovery imaging of the third interstellar object, I3/ATLAS, providing constraints on its composition \citep{feinstein_precovery_2025,martinez-palomera_pre-discovery_2025}.

\subsection{Dwarf Planets and Planet Nine}

K2 Campaign 7 included observations of the Pluto and Charon system, providing a continuous high-precision light curve of the dwarf planet in our solar system \citep{benecchi_k2_2018}. Pluto and Charon were unresolved within a K2 pixel, and a dedicated mask was used to obtain observations along the K2 detectors. The light curve showed clear variability due to the rotation of Pluto and Charon, covering 13 rotation cycles during the campaign.  The data showed a smaller amplitude than predicted, consistent with a model in which Pluto continues to move material on its surface through evaporation/re-deposition processes. The variability was also found to be stable over a 3 month baseline, consistent with the small heliocentric distance change during the timespan of the observations. 

Thanks to its all-sky coverage, TESS data has been suggested to be an effective tool for discovering distant solar system objects, including the hypothesized Planet Nine. By shifting and stacking individual FFI exposures, studies using a few sectors of TESS data have recovered TNOs down to $V\approx22$ and discovered several new candidates \citep{rice_exploring_2020}. This implies sensitivity to an object like the predicted Planet Nine to a distance of about 900\,au \citep{holman_tess_2019, payne_tess_2019}.

\subsection{Ice Giant Planets} 

The ecliptic pointing of the K2 Mission campaigns allowed uninterrupted observations of Neptune and Uranus. A key motivation was to detect intrinsic oscillations, which would provide insights into the interior structure of giant planets \citep{vorontsov_free_1976}, similar to the constraints on Saturn's interior from ring seismology \citep{mankovich_diffuse_2021}. K2 obtained fast-cadence (1-min) observations of Neptune over 49 continuous days \citep{rowe_time-series_2017}. No oscillations were detected, but the data provided an upper limit of $\approx$\,5\,ppm. The data did show remarkable variability on a range of timescales due to clouds in Neptune's atmosphere \citep[][Figure \ref{fig:solar}]{simon_neptunes_2016}, including consistent variability due to a single discrete storm and smaller or fainter clouds that contribute to short-term brightness variability. The observations provide context for spatially unresolved observations of substellar objects with cloud variability, such as directly imaged free-floating brown dwarfs.

A remarkable discovery from K2's Neptune observations was the detection of solar oscillations in reflected light \citep{gaulme_distant_2016}. This detection has important implications since solar reference values for global asteroseismic parameters are frequently used to infer fundamental properties of stars. The frequency of maximum power is expected to depend on wavelength \citep{howe_solar_2020}, but until that point had not been measured in Kepler's bandpass. The analysis of K2 data showed a 5\% higher value than typical adopted values from other solar observations, potentially indicating a bias which would propagate to mass and radius estimates for stars other than the Sun. The difference may also be consistent with the stochastic nature of the excitation of solar-like oscillations.

\section{Citizen Science}

Space-based time-domain missions have helped usher in an era in which non-professional astronomers -- citizen scientists -- have made significant contributions to our understanding of astrophysics. This is predominantly enabled by two developments: the availability of well-documented, open-source software tools which lower the barrier of interacting with data products, and dedicated efforts by the research community to facilitate the analysis of data by citizen scientists. An example of the former is the \verb|lightkurve| software tool \citep{lightkurve_collaboration_lightkurve_2018}, initially developed by the Kepler/K2 science office and now maintained by a growing number of community members. \verb|lightkurve| has been cited over 1000 times in less than 10 years, illustrating its high impact within the community. Many other open-source software tools have been generated to help facilitate interactions with space-based time-domain data, ranging from specific applications (e.g. transit fitting, asteroseismology) to generating light curves from TESS full-frame images. A non-exhaustive list can be found on the TESS Science Support Center webpage (\url{https://heasarc.gsfc.nasa.gov/docs/tess/data-analysis-tools.html}).

\begin{marginnote}[]
\entry{Citizen Science}{Participation of non-professionals in research, enabled by open data policies, open-source software, and community platforms.}
\end{marginnote}

A successful example to engage citizen scientists in the analysis of space-based time-domain data is the Planet Hunters project, launched in 2010 to analyze Kepler data. Planet Hunters used the Zooniverse platform to allow citizen scientists to visually inspect light curves via a web interface. Due to the large amount of light curves produced by Kepler, traditional automated approaches inevitably miss some interesting signal. Citizen scientists discovered new planet candidates \citep{fischer_planet_2012}, including planets in interesting parameter spaces such as a transiting circumbinary planet in a quadruple star system \citep{schwamb_planet_2013}. The most well-known example of a citizen science discovery from Kepler is Boyajian's star \citep[][see Section 4]{boyajian_planet_2016}, a non-standard signal that has since sparked new software tools to automatically look for similar signal of arbitrary shape \citep{wheeler_weird_2019}. Citizen scientists also contributed several discoveries with data from the K2 Mission, including K2-138, a system of five sub-Neptune-sized planets in a near-resonant chain \citep{christiansen_k2-138_2018} and K2-288Bb, a super-Earth-sized planet possibly orbiting the secondary star in a binary system \citep{feinstein_k2-288bb_2019}.

The success of citizen science continued with the TESS Mission, including Planet Hunters TESS \citep{eisner_planet_2020} and Planet Patrol \citep{kostov_planet_2022}. Both projects resulted in large numbers of classifications and used algorithms to combine or refine individual classifications \citep{tardugno_poleo_notplanet_2024}. Notable TESS discoveries include a planet around an evolved subgiant star in a long-period orbit and a multi-planet system orbiting a bright G-type star \citep{eisner_planet_2021}. The examples described above have sparked the community-organized Visual Survey Group, which aims to discover exoplanets using Kepler/K2 and TESS \citep{kristiansen_visual_2022}. The success of online platforms to engage citizen scientists has also expanded beyond exoplanet science: the Zooniverse project ``Cosmic Cataclysms'' aims to vet transient events which are identified using automated machine-learning methods in TESS full-frame image data \citep{roxburgh_tessellate_2025}.

The discovery of transiting planets around bright stars by TESS has also enabled data contributions from amateur astronomers for validating planet candidates. Ground-based, seeing-limited timeseries photometry of stars near TESS planet candidates are now routinely used to rule out eclipsing binaries that may be blended with the target star given TESS's large (21'') pixels. These observations, which are coordinated through the TESS Follow-up Program \citep{collins_tess_2019}, are largely provided by the amateur astronomer community and have been invaluable to prioritize radial velocity follow-up resources. The TESS Follow-Up Program (TFOP) also provided a unique collaboration platform for professional astronomers, including data sharing via the Exoplanet Follow-up Observing Program (ExoFOP) website \citep{christiansen_nasa_2025}. TFOP has enabled mass measurements of hundreds of transiting exoplanets discovered by TESS, many of which are now targets for atmospheric studies with the James Webb Space Telescope.

\section{The Future of Space Based Time-Domain Photometry} 
\label{ch:future}

Several space-based time-domain telescopes are funded to launch before the end of this decade. ESA's PLATO Mission, with a planned launch in late 2026, will obtain fast-cadence, high-precision photometry for approximately one million stars to detect transiting exoplanets and measure asteroseismic ages of the host stars \citep{rauer_plato_2025}. PLATO's field of view will be larger than Kepler's, targeting brighter stars that are more amenable to asteroseismology and radial velocity follow-up. The planned L2 orbit will allow much longer continuous observing campaigns than TESS, thus opening up the discovery space for long-period planets. The current  strategy for PLATO is to observe a field in the southern hemisphere for 2 years \citep{nascimbeni_plato_2025}, with a possible extension of 4 years, or longer with an extended mission. PLATO's observing strategy is thus intermediate to Kepler (narrow and deep) and TESS (shallow and wide).

The Roman Space Telescope, scheduled for launch in late 2026, is NASA's next flagship mission \citep{spergel_wide-field_2015}. While formally a multi-purpose observatory, Roman's large field of view (one hundred times larger than Hubble) will provide powerful time-domain opportunities. The Galactic Bulge Time-Domain Survey, one of three core-community surveys, will obtain 420 days of high-cadence (12 min) photometry in the H-band over a 2 square degree field of view over 5 years along the galactic bulge and in the galactic center \citep{observations_time_allocation_committee_roman_2025}. The primary science goal is the detection of exoplanets using gravitational microlensing \citep{penny_predictions_2019}, but the data will enable a wide range of ancillary science, including asteroseismology of hundreds of thousands of red giants \citep{huber_asteroseismology_2023, weiss_modeling_2025}, the detection of tens of thousands of transiting exoplanet candidates \citep{wilson_transiting_2023}, the detection of stellar-mass black holes \citep{lam_roman_2023}, and probing the variability of stellar populations in the galactic center \citep{terry_galactic_2023}. Additional time-domain surveys may be conducted as guest observer programs. Roman will provide the first high-precision space-based light curves of a large number of stars at infrared wavelengths.

The Earth 2.0 (ET) Mission, with a launch targeted for 2028, aims to detect Earth-analog planets using transits and to probe the demographics of planets on wide orbits using microlensing \citep{ge_et_2022}. The field of view covered by the transit telescopes will include the Kepler field. ET's nominal mission duration is 4 years, thus providing a total of 8 years of continuous baseline for Kepler stars, and over a 20-year total baseline. The microlensing telescope will provide light curves in the galactic bulge, and use simultaneous ground-based observations to obtain masses of planets by measuring microlensing parallaxes. 

Space telescopes that are not dedicated to time-domain photometry also provide powerful datasets. The ESA Gaia mission, in addition to revolutionizing our understanding of the kinematics of stars in our galaxy, has detected over ten million variable stars based on sparsely sampled $\approx$ 50 photometric epochs through DR3 \citep{eyer_gaia_2023, hey_confronting_2024}. The release of Gaia DR4, which includes light curves, is planned for the end of 2026. The James Webb Space Telescope (JWST) currently obtains light curves with the highest photometric precision per unit time ever obtained in the optical/IR as part of its goal to probe exoplanet atmospheres, with datasets showing evidence for stellar variability due to granulation on Sun-like stars \citep{splinter_precise_2025, coulombe_highly_2025} or flares on low-mass stars \citep{howard_characterizing_2023}. Finally, missions dedicated to the targeted follow-up study of exoplanet transits and atmospheres also enable to probe general stellar variability. This includes CHEOPS \citep{benz_cheops_2021}, which launched in 2019, as well as the future planned missions Pandora \citep{quintana_pandora_2024}, Twinkle \citep{stotesbury_twinkle_2022} and Ariel \citep{tinetti_chemical_2018}. 

\begin{figure}
\begin{center}
\includegraphics[width=6in]{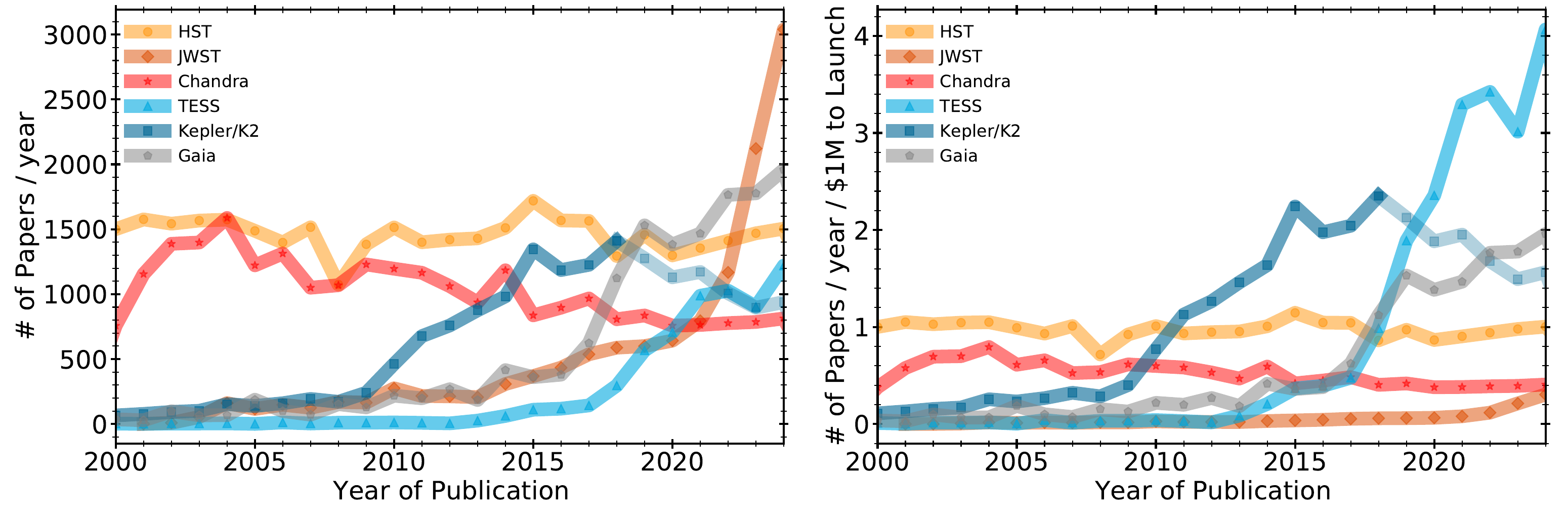}
\caption{\textbf{Space-based time-domain telescopes with open data policies show remarkable scientific productivity.} Left: Number of papers per year using data from flagship missions (HST, Chandra, Gaia, JWST) and dedicated time-domain missions (Kepler/K2 and TESS) over the last 25 years. Numbers were estimated through a NASA/ADS query for mission names appearing in abstracts. An offset of $-200$ papers/year, estimated from pre-2000 publications, was applied for Kepler to account for abstracts unrelated the spacecraft. Kepler/K2 is shown in lighter shading after 2018, when the mission stopped taking data.  Right: Same as left but normalized by the cost to launch in millions of dollars. The following costs were assumed: HST=1.5B, JWST=10B, Chandra=2B, Gaia=1B, Kepler=600M, TESS=300M.}
\label{fig:future}
\end{center}
\end{figure}

Important frontiers in time-domain astrophysics remain to be explored with future missions that are not yet funded. The HAYDN mission concept \citep{miglio_haydn_2021} aims to perform dedicated optical high-precision continuous photometry of stars in crowded fields, such as galactic globular clusters and the galactic bulge, which were inaccessible to previous missions such as Kepler/K2 and TESS due to their large pixel size. Unlike Roman, HAYDN would provide dedicated long-term monitoring in crowded fields, enabling the most precise characterization of stars using asteroseismology and addressing open problems in stellar models that will affect nearly all fields in astrophysics. The Early Evolution Explorer mission \citep[EVE,][]{macgregor_eve_2025} aims to fill the gap of obtaining multicolor space-based photometry to study the exoplanets, activity, and accretion in stars in young star-forming regions. Multicolor photometry will also provide strong benefits for a variety of other science cases, including mode identification in asteroseismology.

Dedicated space-based time-domain telescopes have had an astounding scientific impact in astronomy over the last two decades. Kepler/K2 and TESS data are currently contributing to publications at a rate comparable to NASA's active Great Observatories such as Hubble and Chandra, and exceed the productivity of all major flagship missions when normalized by cost to launch (Figure \ref{fig:future}). This remarkable impact is likely due to the breadth of science enabled by time-domain observations and the open data policy adopted by Kepler/K2 and TESS, which allows scientists from around the world to make scientific discoveries without proprietary data periods. This success highlights that open time-domain missions should remain a cornerstone for the future of space-based astronomy. 

\begin{summary}[SUMMARY POINTS]
\begin{enumerate}
\item Space-based time-domain photometry has transformed astrophysics by providing uninterrupted, high-precision light curves of millions of objects, with impacts across nearly all areas of astronomy.

\item Major advances in stellar astrophysics include the detection of solar-like oscillations in tens of thousands of stars with convective envelopes, the discovery of unexpected behavior in the spin-down of stars, a tight correlation between granulation and oscillations, the realization that the Sun may be an outlier in its activity level compared to other stars, the detection of a wide variety of flares and outbursts across different types of stars, and the discovery of exotic high-order binary systems.

\item Ages measured for stellar populations in the Milky Way have confirmed the basic picture of galactic chemical evolution and radial migration, and provided a chronology for merger events with satellite galaxies. Space-based light curves are also starting to  constrain the demographics of ``dark'' galactic populations such as solar-mass black holes and neutron stars.

\item Extragalactic observations have yielded evidence for new physical processes in supernova explosions from early-time light curves, revealed that the correlation and typical variability timescale for active galaxies extends to low masses, and uncovered variations in the properties of repeating tidal disruption events.

\item Observations in the solar system demonstrated that rotation period distributions (and thus compositions) of asteroid populations are more diverse than previously thought, revealed complex cloud variability in ice giant planets, and yielded new candidates for trans-Neptunian objects in the far reaches of our solar system.

\item Open data policies and open-source software have enabled citizen scientists and amateur astronomers to play key roles in discoveries, ranging from new exoplanets to unusual stellar variability. They also contributed to the remarkable scientific impact of space-based time-domain missions, especially when normalized by mission cost to launch.

\end{enumerate}
\end{summary}

\section*{DISCLOSURE STATEMENT}
The author is not aware of any affiliations, memberships, funding, or financial holdings that
might be perceived as affecting the objectivity of this review. 

\section*{ACKNOWLEDGMENTS}
This review is dedicated to the Principal Investigators, Architects and Engineers of space-based time-domain missions, who paved the way for numerous discoveries and careers in astrophysics. I am grateful to Conny Aerts for the encouragement to write this review, and for comments on the manuscript. I am also grateful to Tim Bedding, Maria Bergemann, Susan Benecchi, Ann Marie Cody, Christina Hedges, Adina Feinstein, Veselin Kostov, Savita Mathur, Armin Rest, Krista Lynne Smith and Josh Winn for helpful comments and suggestions. Finally, I thank Friedrich Anders for providing Figure 9. Funding support is acknowledged from the Alfred P. Sloan Foundation, the National Aeronautics and Space Administration (80NSSC19K0597, 80NSSC21K0652, 80NSSC24K0621, 80NSSC22K0781) and the Australian Research Council (FT200100871). The Kepler and TESS data used in this paper were obtained from the Mikulski Archive for Space Telescopes (MAST) at the Space Telescope Science Institute (STScI) and are freely available via the MAST archive: https://archive.stsci.edu/. MAST is operated by the Association of Universities for Research in Astronomy, Inc., under NASA contract NAS5-26555. Support to MAST for these data is provided by the NASA Office of Space Science via grants NAG5-7584 and NNX09AF08G, and by other grants and contracts. Funding for the Kepler mission was provided by NASA’s Science Mission Directorate, and provided by the NASA Explorer Program for the TESS mission.

%

\bibliographystyle{ar-style2}
\bibliography{zotero}

\end{document}